\documentclass{aa}  

\usepackage{graphicx}
\usepackage{txfonts}
\usepackage{longtable}
\usepackage{subcaption}
\usepackage{float}
\usepackage{adjustbox}

\usepackage{placeins}           % useful with \FloatBarrier, to keep 
\usepackage{orcidlink}          % allow ORCID ids
\usepackage{hyperref}
\usepackage{multirow}

\usepackage{geometry}
\usepackage{supertabular}                                             
\usepackage[table]{xcolor}
\usepackage{overpic}

\usepackage{color}
\definecolor{red(new)}{RGB}{200, 50, 60}

\usepackage{color}
\definecolor{todo}{rgb}{0.89,0.0,0.13}

\usepackage{color}
\definecolor{review}{RGB}{191, 66, 245}

\usepackage{color}
\definecolor{green(new)}{RGB}{50, 200, 60}

\usepackage{color}
\definecolor{cadetgrey}{rgb}{0.57, 0.64, 0.69}

\usepackage{color}
\definecolor{oror}{RGB}{0,150,0}

\newcommand{\comment}[1]{\iffalse{#1}\fi}

\usepackage{amsmath}

\usepackage{graphicx}
\usepackage{txfonts}
\begin{document}

   \title{The mass discrepancy problem in massive stars: an inferential perspective}

\author{C. I. Eze \orcidlink{0000-0003-3119-0399} \inst{\ref{CAMK},\ref{UNN},\ref{KUL}} \fnmsep\thanks{Corresponding author} \and
        G. Handler \orcidlink{0000-0001-7756-1568} \inst{\ref{CAMK}} 
        \and 
        M. Vanrespaille \orcidlink{0000-0002-0420-2473} \inst{\ref{KUL}}
        \and
        D. M. Bowman \orcidlink{0000-0001-7402-3852} \inst{\ref{NCL},\ref{KUL}}
        \and C.\ Aerts \orcidlink{0000-0003-1822-7126} \inst{\ref{KUL},\ref{Radboud},\ref{MPIA}}
       }

\institute{Nicolaus Copernicus Astronomical Centre, Polish Academy of Sciences, Bartycka 18, PL-00-716 Warsaw, Poland \\
           \email{cheze@camk.edu.pl} \label{CAMK} \and
           Department of Physics and Astronomy, Faculty of Physical Sciences, University of Nigeria, Nsukka, Nigeria\label{UNN} 
           \and
           Institute of Astronomy, KU\,Leuven, Celestijnenlaan 200D, 3001 Leuven, Belgium \label{KUL}
           \and
           School of Mathematics, Statistics and Physics, Newcastle University, Newcastle upon Tyne, NE1 7RU, UK \label{NCL} \and
           Department of Astrophysics, IMAPP, Radboud University Nijmegen, PO Box 9010, 6500 GL Nijmegen, The Netherlands\label{Radboud}\and
           Max-Planck-Institut für Astronomie, Königstuhl 17, 69117 Heidelberg, Germany\label{MPIA}
           }

 % \date{Received September 15, 1996; accepted March 16, 1997}
% \abstract{}{}{}{}{} 
% 5 {} token are mandatory
 
  \abstract
   {Bayesian inference has become a standard tool for inferring stellar parameters from stellar evolution models. However, comparatively little attention has been paid to how alternative Bayesian likelihood formulations influence inferred stellar masses, convective-core masses, mixing parameters, and the resulting evolutionary mass discrepancy for massive main sequence stars.}
   {We investigate how alternative Bayesian likelihood formulations influence the inferred stellar parameters and the resulting evolutionary mass discrepancy, while examining the impact of different mixing levels near the convective core and in the radiative envelope on these inferences. }
   {We analysed a sample of seven detached double-lined eclipsing B type binaries with $\beta$~Cep components using one pre-computed MESA grid of stellar evolution models and observables such as mass, effective temperature, surface gravity and luminosity derived from the eclipsing binary modelling. Stellar parameters were inferred within a Bayesian inference method employing nested sampling and piecewise cubic interpolation across the model grid. We compared several inference strategies, including single-star and coeval binary models, with and without dynamical masses included in the likelihood function. Bayesian evidence and Pareto-optimal model selection were used to evaluate the preferred solutions.}
   {Our work reinforces that core-boundary mixing is required for all stars. Further,
   the inferential analysis demonstrates that different likelihood formulations may lead to systematically different inferred mass discrepancies. For the adopted input physics and the small sample of stars, we find that about half
%   Thus, a substantial fraction (about 55-60\%) 
of the apparent reduction in the inferred mass discrepancy may arise from the construction of the Bayesian likelihood, while enhanced near-core mixing accounts for the remaining improvement. A comparison of the alternative likelihood formulations shows that Bayesian evidence alone does not always identify the models with the smallest evolutionary mass discrepancy, motivating the use of Pareto-based model selection to identify the preferred solutions.  The preferred coeval solutions further indicate that the systems span central hydrogen mass fraction of \(0.632\) -- \(0.205\), with ages of \(4.25\) -- \(12.27\) $\rm Myr$ determined with precisions of \(2\) -- \(13\%\) under the adopted model input physics.}
   {The inferred evolutionary mass discrepancy deduced from the used stellar model grid was found to depend on both the adopted mixing prescription and the formulation of the Bayesian likelihood. Comparing alternative likelihood formulations helps to distinguish methodological effects from physical contributions. }


   \maketitle
%
%________________________________________________________________

\section{Introduction} 
\label{sec:introduction}
The determination of stellar masses is fundamental to stellar astrophysics. Stellar masses can be inferred dynamically through spectroscopic and eclipsing binary observations or derived from stellar evolution models --- see review by \citet{Serenelli2021}. The resultant spectroscopic, and in the case of binary systems also the dynamical masses, are typically discrepant with the inferred masses from the evolutionary models, resulting in the mass discrepancy problem \citep{Herreroetal1992}. 
In massive stars, this discrepancy has been partially attributed to the underestimation of the convective core mass because of unconstrained convective-boundary mixing (CBM; \citealt{Pedersenetal2021,AndersandPedersen2023}) processes, in addition to limitations in atmospheric modelling \citep[e.g.][]{Tkachenkoetal2020}. Since massive stars have convective cores and radiative envelopes during the main sequence, CBM can increase the convective core mass and prolong their lifetimes (see \citealt{Johnston2021b}). Enhanced CBM has been reported to decrease the mass discrepancy in massive binary systems \citep{Masseyetal2012, Tkachenkoetal2020}. 

Moreover, accounting for other mechanisms, such as rotation 
 \citep{Masseyetal2012, Markovaetal2018}, has been proposed to reduce the mass discrepancy. \citet{WeidnerandVink2010} argued that the mass discrepancy may yet be solved from a proper calibration of all observed parameters, stimulating a debate on whether the discrepancy reflects the need for improved physics in structure and evolution models or better treatment of measurements
 and modelling tools. Therefore, accurate measurement of the convective core masses and the efficiency of interior mixing processes of massive stars may hold the key to resolving the mass discrepancy problem.

\begin{table*}[ht]
\begin{center}
\caption{Stellar sample of double-lined spectroscopic eclipsing binaries with a $\beta$ Cep component and their reported observed parameters in the literature. The * symbol denotes that a fixed value of the effective temperature from the referenced study was used during binary modelling. For such cases, we adopted an uncertainty of 480 K, which is the average reported uncertainty of the $T_{\rm eff}$ of the other primary components' uncertainties in the sample, as the observed uncertainty during the Bayesian modelling. The labels of A and B denote the primary and secondary components, respectively, with the primary being the more massive one.}\label{tab:sample_parameters}
%\begin{adjustbox}{angle=90}
\begin{tabular}[t]{lcccccc}
\hline\hline
Name &  Component  &  $M \ (\mathrm{M}_\odot)$ 	   &	$T_\mathrm{eff} \ (K)$	&	$\log L \ (\mathrm{L}_\odot)$	&$\log g \ ($cm$\,$s$^{-2}$) &	Reference\\
\hline
%CPD-41 7746 & A & 17.25(3) & 31100* & 4.46(3) &4.11(2) & \citet{Cakirlietal2025a}\\
CW Cephei & A& 12.951(52) & 28300(460) & 4.243(29) & 4.066(6) & \multirow{2}{*}{\citet{LeeandHong2021}} \\
CW Cephei & B& 11.877(49) & 27431(434)& 4.119(28) & 4.099(6)& \\
HD 254346 & A & 15.52(4) & 28000* & 4.57(4) & 3.79(3) & \multirow{2}{*}{\citet{Cakirlietal2025a}}\\
HD 254346 & B & 10.77(5) & 16300(420) & 3.48(5) & 3.79(3)\\
HD 74455 & A& 16.30(3) & 29700* & 4.36(4)  & 4.08(3) & \multirow{2}{*}{\citet{Cakirlietal2025a}}    \\
HD 74455 & B &  10.32(4) & 23300(450) & 3.79(4) & 4.14(4)    \\
KX Vel & A & 13.68(3) & 26000* & 4.451(35) & 3.738(32) & \multirow{2}{*}{\citet{Cakirlietal2025B}}   \\
KX Vel & B & 8.28(4) & 22100(220) & 3.416(59) & 4.272(41)     \\
$\varphi$ Ori & A &  17.69(5) & 25600* & 3.59(3) & 4.06(3) & \multirow{2}{*}{\citet{Cakirlietal2025a}} \\
$\varphi$ Ori & B & 11.61(5) & 19900(420) & 3.99(4) &  4.28(4)    \\
V453 Cygni & A & 13.96(23) & 28800(500)& 4.666(31) & 3.708(18) & \multirow{2}{*}{\citet{Southworthetal2020}}   \\
V453 Cygni & B & 11.10(18) & 27700(600) & 4.163(39) & 4.044(9)    \\
V1295 Sco & A & 17.26(2) & 30500* & 4.724(25) & 3.843(7) & \multirow{2}{*}{\citet{Cakirlietal2025B}}     \\
V1295 Sco & B & 12.66(3) & 28100(390) & 4.227(33) & 4.064(9)      \\
%VV Ori & A & 9.52(56) & 26200(1000) & 4.02(7) & 4.026(11) & \citet{Southworthetal2021}     \\

\hline\hline
\end{tabular}
%\end{adjustbox}
\end{center}
\end{table*}

Eclipsing binaries with pulsating components are among the best candidates for the study of the mass discrepancy, as they offer the opportunity to apply either binary dynamical constraints or asteroseismic constraints or both --- see review by \citet{SouthworthandBowman2025}. Masses from eclipsing binary modelling can be measured to an excellent accuracy
%precision 
of better than 1\% \citep{Southworth2012, Maxted2020a, Serenelli2021}. Hence, such systems are benchmarks for measuring the mixing efficiency and offer great potential to resolve the mass discrepancy. For example, eclipsing binary systems have been used to help calibrate CBM in intermediate-mass stars \citep{ClaretandTorres2016, ClaretandTorres2017, ClaretandTorres2018, ClaretandTorres2019}, and demonstrate that CBM increases with mass up from about $1 \, \mathrm{M}_{\odot}$ to about $2 \, \mathrm{M}_{\odot}$ and then flattens,
as also found from pressure-mode asteroseismology of single stars \citep{Angelou2020}.
\citet{Pedersen2022} studied the core masses for the range $[3,9]\,$M$_\odot$ from gravito-inertial  modes and found an increasing trend from their sample of 26 Slowly Pulsating B (SPB) stars.
\citet{ConstantinoandBaraffe2018} revisited eclipsing binaries in the mass range of $1.25 - 4.2\, \mathrm{M}_{\odot}$ and could not recover the previously reported stellar mass dependence. They discussed that there is substantial uncertainty in the derived mass dependence owing to model degeneracies, and suggested that the errors in previously reported CBM trends in mass could be 
%drastically 
underestimated. Moreover, the mass of the convective core is strongly dependent on the evolutionary stage, as illustrated from gravity-mode asteroseismology \citep{Mombarg2019,Mombarg2021}. To tackle the effect of mass on the value of CBM properly, one must thus use methods that are capable of high-precision age-dating via estimation of the central hydrogen mass fraction still left in the burning convective core.

The mass dependence of CBM has also been reported in the literature  for stars at the high mass end \citep[e.g.][]{Ribasetal2000}, with larger CBM having been reported to partially resolve the mass discrepancy. Using a sample of 21 O-type stars in eclipsing binaries with masses $14-56$~M$_{\odot}$, including those in the Large and Small Magellanic Clouds, \citet{WeidnerandVink2010} obtained good agreement between evolutionary model and dynamical masses, suggesting that the discrepancy might have been resolved. However, \citet{Masseyetal2012} studied two O-type eclipsing binaries in the Large Magellanic Cloud and found them to be under-massive by 11\% or over-luminous by 0.2~dex. The apparent differences observed in the reported possible agreement (or disagreement) of the dynamical masses with the evolutionary masses could be reflecting possible systematics or methods in constraining the mixing parameters  rather than only the missing physics not accounted for by the mixing parameters.  In particular, \citet{Tkachenkoetal2020} conducted a detailed analysis of CBM to reproduce the dynamical masses of 11 eclipsing binary stars with masses of $4.58-17.07$~M$_{\odot}$ derived from binary modelling based on TESS data. They concluded that enhanced CBM partly resolves the mass discrepancy at a given metallicity, while also showing that a change in metallicity significantly affects the mass of a star and its position in the HR diagram. Further observational mapping of the mass discrepancy is thus warranted (e.g. \citealt{Tkachenkoetal2024}).

Previous studies have largely focused on improving stellar evolutionary physics through enhanced CBM, rotation effects, or additional mixing processes for many reasons \citep{AT2024}, including
to reduce the evolutionary mass discrepancy. Comparatively less attention has been paid to the influence of the statistical inference approach itself. In particular, when dynamical masses are incorporated directly into Bayesian likelihood functions, their exceptionally small uncertainties may dominate the inference, making it difficult to distinguish genuine improvements arising from stellar physics from those introduced by the construction of the likelihood. The primary objective of this work is to investigate the mass discrepancy using multiple Bayesian inferential formulations designed to separate methodological effects from physical effects of enhanced CBM to the inferred mass discrepancy, while simultaneously deriving the core properties such as the convective core mass (M$_{\rm cc}$), central hydrogen mass fraction (X$_{\rm c}$) and mixing parameters. To this end, we formulate several complementary likelihood constructions, introduce Bayesian multi-objective model selection \citep[e.g.][]{miettinen1999, ehrgott2000, Deb2001} to balance statistical fit quality against mass-discrepancy reduction, and compare inference under both single-star and coeval binary evolutionary models. The resulting inferential strategies enable us to derive the convective core mass and mixing parameters, as well as to quantify how much of the apparent reduction in the mass discrepancy originates from improved mixing parameters and how much reflects the inferential methodology itself. 

We describe the sample in Section~\ref{sec:sample_description}, and the evolutionary models in Section~\ref{sec:stellar_models}. The analysis framework is described in Section~\ref{sec:analysis}. In Section~\ref{sec:results} we describe the results, and we discuss the results and conclude in  Section~\ref{sec:discussion}.
%, and conclude in Section~\ref{sec:conclusions}.

\section{Sample Description} 
\label{sec:sample_description}
%Double-lined spectroscopic (SB2) eclipsing binaries enable a precise determination of the masses of both components.  
To evaluate the mass discrepancy, we select a small sample of double-lined spectroscopic (i.e. SB2) eclipsing binaries of spectral type B, which contain at least one $\beta$~Cep component. We specifically focus on binaries with a $\beta$~Cep component as these are common among massive stars (i.e. $M \gtrsim 8$~M$_{\odot}$; \citealt{Burssensetal2020, 2026MNRAS.tmp.1395T}).
The $\beta$~Cep stars pulsate in low-order pressure or gravity modes that allow for robust asteroseismic inferences \citep{Bowman2020,Burssensetal2023,Fritzewskietal2024,Vanlaer2025}. 
%Although our sample contains $\beta$~Cep stars, which are good laboratories for asteroseismic experiments (see \citealt{Bowman2020}), 
In the current study, we do not exploit any asteroseismic constraints but limit ourselves to the
%are not yet included in this analysis, which forms the first part of the experiment. Thus, only 
binary observables. 
%Stellar evolutionary models are used 
This has the advantage that our current work can be applied to non-pulsating binaries as well. 
We leave the inclusion of stricter additional inferences from
identified pulsation modes to future work.

To ensure homogeneity and consistency, only systems with complete published parameters are considered. We assembled a sample of seven such pulsating eclipsing SB2 systems, which have component masses that range between $8-20$~M$_\odot$.  Consequently, we have a total of 14 stars in the sample. Table \ref{tab:sample_parameters} shows the list of systems in our sample, which includes their reported dynamical masses, effective temperature, luminosity, and surface gravity. Any system missing any of these four observational parameters is considered to have incomplete parameters and is not included in the sample. 

We note that CW~Cephei and V453~Cygni have been previously modelled in specific studies of 
the mass discrepancy using a $\chi^{2}$ merit function \citep{Tkachenkoetal2020}. Our larger sample and more sophisticated statistical inference treatment in the covered mass regime are
the major novelties of this work.

\section{Stellar Models}
\label{sec:stellar_models}
In stellar evolution models, the total time evolution of the  mass fraction $X_{i}$ of a chemical element $i$ is given as \citep{Kippenhahn1990, AndersandPedersen2023}: 

\begin{align}
\frac{\partial X_i}{\partial t} 
&= \frac{m_{i}}{\rho} \left[\sum_{j} R_{ji} -\sum_{k} R_{ik} \right]  + \frac{1}{\rho r^2} \frac{\partial}{\partial r} \left[ \rho r^2 D_{\text{mix}} \frac{\partial X_i}{\partial r} \right] + \mathcal{M}_i^{\text{micro}}, \label{Eq:mixing_overall}
\end{align}

\noindent where $r$ is the radius coordinate, $m_{i}$ is the mass of the $i^{\rm th}$
chemical element, R$_{\rm ji}$ and R$_{\rm ik}$ denote the nuclear reaction rates producing and destroying species $i$, respectively, $\rho$ is the local density, $D_\mathrm{mix}$ is the turbulent diffusivity  in units of $\text{cm}^2\,\text{s}^{-1}$ 
\citep{Freytagetal1996, AndersandPedersen2023},
and $\mathcal{M}_i^{\text{micro}}$ accounts for extra microscopic atomic diffusion effects 
such as gravitational settling and radiative levitation (see \citealt{Mombarg2020,Mombarg2022,AlecianDeal2023}). 

The term $D_\mathrm{mix}$ in 
Eqn.~(\ref{Eq:mixing_overall}) 
assumes that CBM can be described diffusively and
is used to capture the combined contributions from different physical processes such as convection, magnetic fields, rotation, and waves in different regions of the star that are not included otherwise.  Hence for a slowly rotating, non-magnetic massive star, the main contribution to $D_\mathrm{mix}$ is from the convective core, the CBM region, and mixing in the radiative envelope. Therefore, $D_\mathrm{mix}$ is defined following \citet{AndersandPedersen2023}, such that: 

\begin{equation}
D_{\text{mix}}(r) = D_{\text{conv}}(r) + D_{\text{CBM}}(r) + D_{\text{env}}(r), \label{Eq:mixing}
\end{equation}

\noindent where $D_\mathrm{CBM}(r)$ accounts for the mixing in the convective boundary region. In fast rotating and/or magnetic stars, the mixing contributions from rotation and/or magnetic fields are usually added to Eqn.~(\ref{Eq:mixing}) as extra terms. In the case of close binary systems, where binary interaction is important, the effect of tides on mixing could also be captured in $D_\mathrm{mix}(r)$ as an additional term.

There are generally two commonly used prescriptions for diffusive CBM: 
an exponentially decaying coefficient or a constant one described by a step function. In the case of an exponential description, the mixing efficiency decays exponentially with distance from the convective boundary, with the diffusion profile \citep{Freytagetal1996, Herwig2000, AndersandPedersen2023} defined as: 
\begin{equation}
D_{\text{CBM}}(r) = D_0 \exp \left[ \frac{-2(r - r_0)}{f_{\text{CBM}} H_{p,0}} \right]\, ,
\end{equation}
\noindent where $f_\mathrm{CBM}$ is the exponential CBM scaling parameter, $D_{0}$ is the convective mixing efficiency at $r_{0}$ and $H_{p,0}$ is the pressure scale height at $r_{0}$.
 At $r_{\rm cc}$, where the convective boundary occurs, the mixing length theory (MLT; \citealt{BohmVitense1958}) predicts the convective velocities of rising convective parcels of gas, and hence the associated convective mixing efficiency, to vanish. The mixing length, $l_\mathrm{mix}$, which is the distance a convective fluid element travels before mixing with the surrounding, is typically implemented as a product of the local pressure scale height, $H_{p}$, resulting in $l_\mathrm{mix}=\alpha_{\rm MLT}H_{p}$ \citep{DengandXiong2001}, such that $\alpha_{\rm MLT}$ is the mixing length parameter. Hence, the exponential CBM is calibrated at
$r_0 = r_{\text{cc}} - f_0 H_{p,\text{cc}}$ where $H_{p,\text{cc}}$ is the pressure scale height at the convective boundary and $f_{0}$, in the case of \textsc{MESA} models, is fixed to values between $0.002-0.005$ \citep{AndersandPedersen2023}. 
The temperature gradient in the overshoot zone is assumed to be the radiative one, given as:
\begin{equation}
\nabla_T = \nabla_{\text{rad}}= \frac{3 \kappa L P}{16 \pi a c G m T^4}. \label{Eq:exponential_overshoot}
\end{equation}
We use this description of convective overshooting in our work.

To determine the properties of the seven pulsating SB2 binary systems in this work, we used precomputed grids of stellar evolutionary models. We used the specific \textsc{MESA} grid calculated by \citet{Vanrespailleetal2026}, which was computed based on a revised version of \citet{Burssensetal2023} using the Modules for Experiment in Stellar Astrophysics \citep[MESA:][]{Paxtonetal2011, Paxtonetal2013, Paxtonetal2015, Paxtonetal2018, Paxtonetal2019, Jermynetal2023}. The varied parameters include mass, central hydrogen mass fraction (X$_{\rm c}$), CBM using a diffusive exponential overshooting ($f_{\rm CBM}$), additional diffusive mixing in the envelope defined as $D_{\rm mix} = D_{\rm mix,0} \frac{\rho_0}{\rho}$ according to the hydrodynamical simulations by \citet{RogersandMcElwaine2017}, where $\rho$ the local density and $\rho_0$ the density at the base of the envelope. A solar metallicity of $Z=0.014$ was used. The mass ranges from 7 to 30~M$_{\odot}$ in logarithmic steps of $\Delta\log \rm M = 0.015$, $f_{\rm CBM}$ from 0.005 to 0.035 in steps of 0.005, $\log D_{\rm mix,0}$ from 1 to 5 (or 2 to 6 when $M > 13\,$M$_\odot$) in steps of 1, and $X_{\rm c}$ from 0.7 to 0.0001 with the resolution increasing as the $X_{\rm c}$ decreases to ensure the end of the main sequence is well sampled. 

In the rest of this paper, we only use this stellar model grid, which was designed specifically for the application of $\beta\,$Cep sample asteroseismology by \citet{Vanrespailleetal2026}. This implies that conclusions are only valid for this particular choice of the input physics that went into this grid. Comparisons among grids of models with a variety of input physics in terms of fit quality, as in \citet{Moravveji2016} or \citet{Pedersenetal2021} are not treated in our work. 

\section{Analysis}
\label{sec:analysis}
Each of the 14 stars in our sample of seven pulsating SB2 eclipsing binaries is assumed to have negligible tidal deformation. First, all stars are treated as effectively single stars, such that they are assumed to have evolved independently of their companion. As a second case, we treat both components in each binary as coeval. Comparison of these two cases provides insight of how the mass discrepancy depends on the assumption of independent and coeval evolution for both components of the binaries.

\subsection{Interpolation framework}
The grid of {\sc MESA} evolutionary models is discrete. As a result, to evaluate the models within the grid as if they were continuous functions, an interpolation scheme was applied. We use a two-step piecewise cubic hermite interpolating polynomial (PCHIP) \citep{Fritsch1982}; first along the evolutionary state ($X_{\rm c}$) in single-star case or age in coeval case, and second across models for fixed values of $f_{\rm CBM}$ and $\log D_{\rm mix,0}$. PCHIP preserves monotonicity and avoids spurious oscillations between grid points \citep{Cox2012}. This is particularly relevant for the chosen evolutionary coordinates (i.e.\ $X_{\rm c}$ and age), which evolve monotonically during a star's main-sequence within the main sequence lifetime. Hence, its use preserves the local shape of the evolutionary tracks, ensures numerical stability of likelihood evaluations, and prevents interpolation-induced artefacts from influencing the statistical results. 

%\subsubsection{Single-star case}
In the single-star case, the interpolation was performed using $X_{\rm c}$ as the evolutionary coordinate. The tracks were sorted in decreasing $X_{\rm c}$ and PCHIP interpolators were built along each track for the parameters $\log T_{\rm eff}$, $\log L$, $\log g$, age, and $M_{\rm cc}$, after removing possible duplicate values of $X_{\rm c}$. 
%Here, $M_{\rm cc}$ denotes the convective core mass and other quantities have their usual meaning. 
%Thus, each available track returns an interpolated stellar model corresponding to the same evolutionary state for a proposed value of $X_{\rm c}$. The resulting quantities were then subsequently interpolated across stellar mass using PCHIP interpolation for fixed $\log D_{\rm mix,0}$ and $f_{\rm ov}$. 

%\subsubsection{Coeval binary case}
In the coeval binary case, the components of a binary system are assumed to  have the same age (and metallicity) since they formed from the same birth cloud. To enforce this coeval behaviour, the interpolation procedure was modified and the stellar age instead of $X_{\rm c}$, was adopted as the evolutionary coordinate. The models were sorted by age and PCHIP interpolators were constructed as functions of age for the parameters $\log T_{\rm eff}$, $\log L$, $\log g$, $X_{\rm c}$, and $M_{\rm cc}$ after removing duplicate values of age. For a proposed common binary age, each track was evaluated at that age. The resulting quantities were then interpolated across stellar mass using PCHIP interpolation, for fixed $\log D_{\rm mix,0}$ and $f_{\rm CBM}$.

\subsection{Statistical Framework} \label{sec:stats-framework}
Using observational constraints derived from the eclipsing binary models, we conducted Bayesian inferential analysis to determine mixing parameter values that best fit the observables and possibly resolve the mass discrepancy. Because covariance matrices for the published binary observables are unavailable, we adopt the standard approximation of independent Gaussian uncertainties for the observables and a simplified Gaussian likelihood function. Thus, the likelihood function, $\mathcal{L}$ based on $\chi^2$ follows 
\begin{equation}
 \mathcal{L}(D \mid \theta, \mathcal{M})
=
\exp\left(
-\frac{1}{2}\chi^2
\right),   
\end{equation}
\noindent where \(D\) are the observed data, \(\theta\) represent the model parameters, \(\mathcal{M}\) is the adopted stellar evolutionary hypothesis (e.g. single-star evolution, coeval binary evolution, or a specific mixing parameter), and 
\begin{equation}
\chi^2 =  \sum_i \left(\frac{O_i - M_i}{\sigma_i}\right)^2\/, 
\end{equation}
with $O_i, \sigma_i$ and $M_i$ the observed quantity, its error, and the theoretical prediction for the quantity, respectively. To estimate the free parameters, 
we use a log-likelihood merit function,
\begin{equation}
\ln \mathcal{L} = -\frac{1}{2}\chi^2 ~ .
\end{equation}
\noindent Using Bayes' theorem, the posterior probability distribution is thus defined as:
\begin{equation}
P(\theta \mid D, \mathcal{M})
=
\frac{
\mathcal{L}(D \mid \theta, \mathcal{M}) \cdot
\,
P(\theta \mid \mathcal{M})
}{
\mathcal{Z}(D \mid \mathcal{M})
}.
\end{equation}
\noindent where $P(\theta \mid \mathcal{M})$ is the prior and $\mathcal{Z}(D \mid \mathcal{M})$ is the posterior evidence. Uniform priors were adopted for the sampled parameters (mass, X$_{\rm c}$, age) over the ranges explored in the model grid. The mass and $X_{\rm c}$ are sampled in the case of single-star models and the mass and age are sampled in the case of coeval binary models, while $f_{\rm CBM}$ and $\log D_{\rm mix,0}$ are treated as discrete hyper-models\footnote{Here, a hyper-model denotes a discrete stellar-model configuration
defined by a prescribed combination of $f_{\rm CBM}$ and $\log D_{\rm mix,0}$, rather than parameters sampled continuously within an inference run.} explored by outer loops in both cases. Binary observables such as $T_{\rm eff}$, $\log g$, $L$, and mass are considered in the likelihood in a controlled experimental manner described in the likelihood framework given in Section \ref{sec:likelihood_framework}. The analysis was conducted through nested sampling \citep[e.g.][]{Skilling2009} with \textsc{dynesty}\footnote{https://github.com/joshspeagle/dynesty.} \citep{Speagle2020, Koposovetal2026}, which is a \textsc{Python} package for estimating Bayesian posteriors and evidences using nested sampling.

For each combination of mixing parameters ($f_{\rm CBM}$, $\log D_{\rm mix,0}$), nested sampling yields the posterior distribution of the sampled stellar parameters together with the Bayesian evidence, $\ln{\mathcal{Z}}$. The posterior distributions are used to infer stellar properties such as the stellar age, M$_{\rm cc}$, evolutionary mass, and the mass discrepancy ($\Delta M$) between the mass of the evolutionary model and the dynamical mass, while $\ln{\mathcal{Z}}$ provides a quantitative measure of the statistical support for each set of mixing parameters after marginalising over the sampled space of the other free parameters. The models are then compared with each other to identify the most likely one based on their Bayesian evidence.

To ensure numerical convergence of the nested-sampling procedure, preliminary calculations were performed with live points ($N_{\rm live}$)\footnote{The number of live points, $N_{\rm live}$, specifies the number of active samples maintained by the nested-sampling algorithm as it progressively explores higher-likelihood regions of the parameter space.} of 150, 300, 350, and 500, together with multiple independent random seeds. Convergence was assessed from the stability of both the Bayesian evidence and the posterior-derived quantities relevant to the mass discrepancy analysis. Evidence estimates obtained at successive $N_{\rm live}$  values were consistent within their nested-sampling uncertainties, while the posterior medians and credible intervals showed no systematic changes with increasing $N_{\rm live}$. Therefore, increasing the number of live points beyond 
$N_{\rm live}=300$ (or 500 for the more complex model formulations) produced no material change in the inferred parameters or model-comparison conclusions. We consequently adopted the lowest $N_{\rm live}$  of 300 (or 500 for the coeval case) for which these quantities were numerically stable, thereby avoiding additional computational expense without a corresponding material gain in inference precision.

\subsection{Likelihood framework}\label{sec:likelihood_framework}
Using the description in Section~\ref{sec:stats-framework}, we built seven different models for each star. These models are defined based on the quantities that are included in the likelihood function, the stellar case, and the parameters used for statistical inferences. The different statistical formulations described below represent different statistical experiments because they are based on different sets of observables and conditions. Consequently, Bayesian evidences are compared only among models evaluated under the same likelihood function definition. For example, the reference  model (RM) and Pareto model (PM) use only $T_{\rm eff}$, $\log \rm L$ and $\log \rm g$ in the likelihood, whereas the mass model (MM), Pareto mass model (PMM) and coeval cases additionally include the dynamical mass. Since these models condition on different data vectors, the values of their evidence ($\ln \mathcal Z$) are not interpreted as directly comparable across the different likelihood frameworks. Instead, the evidence is used to rank competing mixing parameter values within each likelihood definition.

\noindent
\paragraph{\textbf{Reference model (RM):}}
The reference model is the model with minimal mixing ($f_{\rm CBM}= 0.005$ and $\log D_{\rm mix,0} =1.0$). $T_{\rm eff}$, $\log g$ and $L$ are included in the likelihood function to derive the reference model. This gives the baseline evolutionary mass at minimal mixing and a supposedly  underestimated core mass with maximum possible discrepancy.

\paragraph{\textbf{Mass model (MM):}} In this model, the dynamical mass is included in the likelihood function in addition to those in the reference model. Due to the 
accurate measurement of the dynamical masses, the relative uncertainties are usually very small compared to those of other measured quantities. As a result, the likelihood is usually dominated by the mass
%While including mass in the likelihood better constrains the model than when it is excluded, the possible bias towards the dynamical mass 
%poses a problem and 
%sometimes gives a false positive impression of resolving the discrepancy. Because the likelihood is dominated by the mass, 
and the model with the least $\Delta M$ most often gives the maximum evidence in this case. 
%As a result, we adopt the model with the maximum evidence. 

\paragraph{\textbf{Pareto model (PM):}} Here, we remove the dynamical mass from the likelihood similar to the RM, but compute models for all the possible mixing parameter values allowed in the grid. Instead of picking the model with the maximum posterior evidence, we treat the selection of the preferred mixing parameter values as a Bayesian multi-objective model-selection problem. For each combination of the mixing parameters, the nested-sampling analysis yields the Bayesian evidence, $\ln \mathcal Z$, which measures the overall statistical support of the stellar model given the observables included in the likelihood. In addition, the posterior samples are used to evaluate the probability that the inferred evolutionary masses of the stellar components are simultaneously consistent with their independently measured dynamical masses within their combined observational and model uncertainties. We denote this quantity by $P_{\rm consistent}$.

The Bayesian evidence and $P_{\rm consistent}$ quantify complementary aspects of model quality. In simpler terms, the evidence measures the statistical agreement between the stellar evolutionary model and the observational constraints, whereas $P_{\rm consistent}$
measures the probability that the model successfully reproduces the dynamical masses of the components. Since the model that fits the observables may not always be the one that reduces the mass discrepancy, we formulate the model-selection problem as a Bayesian multi-objective Pareto-based optimization \citep[e.g.][]{miettinen1999, ehrgott2000, Deb2001}.

Since the dynamical masses are excluded from the likelihood function, they play no role in determining either the posterior distribution or the Bayesian evidence. Consequently, the inferred evolutionary masses constitute genuine posterior-predictive quantities, and their agreement with the independently measured dynamical masses provides an external test of the adopted mixing parameters. The quantity $P_{\rm consistent}$ therefore measures the predictive success of each mixing parameter in this case, rather than the quality of a fitted observable.

\begin{figure}%[ht]
\centering
\includegraphics[width=0.90\linewidth]{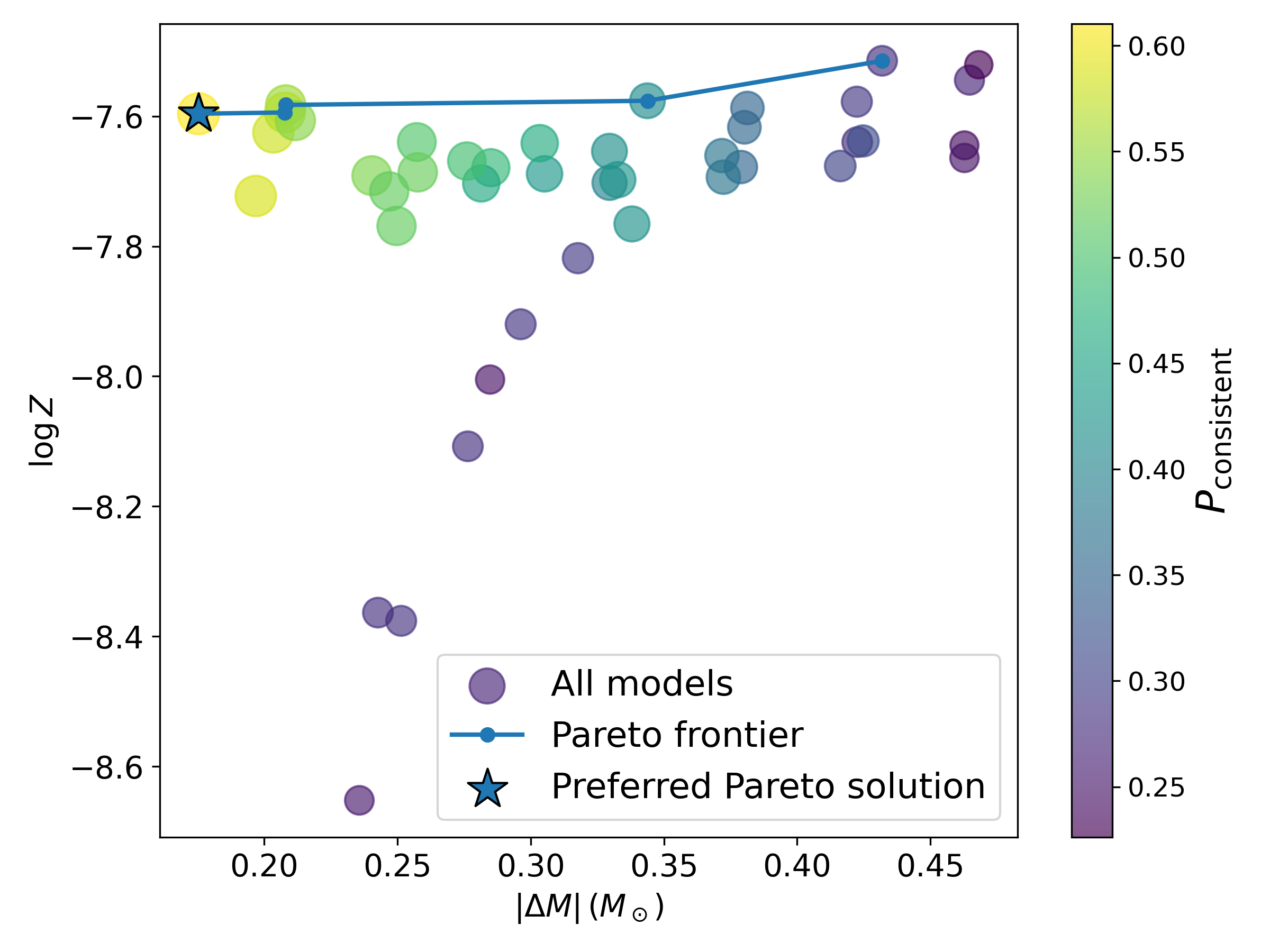}
\caption{Example of the Pareto frontier for CW~Cep. The Bayesian evidence is plotted against the mass discrepancy and colour-coded by the probability that the solution is consistent with zero discrepancy. The thick line is the Pareto frontier connecting the Pareto optimal solutions. 
}
\label{fig:pareto}
\end{figure}

A model is considered Pareto optimal if no alternative model provides concurrently both stronger statistical support, as demonstrated by the evidence, and better agreement with the dynamical mass \citep[see e.g.][]{miettinen1999, ehrgott2000, Deb2001}. The collection of such models forms the Pareto frontier and represents the set of best compromise solutions between the overall statistical fit quality and simultaneous reproduction of the binary masses. 
Figure~\ref{fig:pareto} illustrates this Pareto frontier in the two-dimensional objective space ($P_{\rm consistent}$, $\ln \rm \mathcal{Z}$) for the binary CW~Cep. Each point represents one set of mixing parameter values explored by the Bayesian analysis.

When a single representative solution is required for presentation or subsequent diagnostic analyses in systems where the Pareto-optimal models are more than one, we adopt a Bayesian decision rule. We first retain all Pareto-optimal models whose evidences satisfy
$\Delta \ln \mathcal{Z} = \ln \mathcal{Z}_{\rm max} - \ln \mathcal{Z} \leq 1$
which correspond to models with comparable statistical support. Among these statistically competitive Pareto-optimal models, we select the solution with the largest value of $P_{\rm consistent}$. This procedure avoids introducing an arbitrary scalar merit function or weighting between competing objectives while yielding a representative model for subsequent analysis.

The component-wise mass discrepancies, $\Delta M_{1}$  and $\Delta M_{2}$, are reported separately throughout this work as diagnostic quantities describing the residual evolutionary–dynamical mass offsets of the individual stellar components, but they are not themselves used as optimization objectives. In summary, the PM formulation separates model fitting from model prediction: the atmospheric observables constrain the stellar models through the likelihood function, whereas the dynamical masses provide an independent posterior-predictive assessment of the inferred mixing parameter values.

\paragraph{\textbf{Pareto mass model (PMM):}} The PMM  formulation follows the same multi-objective optimisation strategy as the PM model but differs only in the inclusion of the dynamical masses in the likelihood function. Consequently, the Bayesian evidence reflects the statistical support for the complete set of fitted observables, while the residual mass discrepancy is interpreted as the remaining mismatch after fitting the dynamical masses rather than as an independent model prediction. The Pareto-based procedure is therefore used to balance the overall statistical support for the fitted model with the remaining residual mass discrepancy, thereby distinguishing models that achieve similar statistical support but differ in their ability to minimise the fitted mass residual. 

Models RM, MM, PM, and PMM are computed for the single star case for each of the targets in the sample. However, the components of eclipsing binaries are reported to be coeval.  As a result, enforcing coeval age for both components  adds additional constraints and better captures the evolution of the components of eclipsing binaries.  In subsequent models built for the targets, the coeval age was enforced and the PMM strategy was adopted.

\paragraph{\textbf{Coeval fixed mixing model $\text{(Coeval\_FM)}$:}} Here, we aim to test whether a given set of mixing parameter values can simultaneously reproduce both components of a given binary system in our sample. We sample the masses of both components and the coeval age. In this model, a set of $f_{\rm CBM}$ and $\log D_{\rm mix,0}$ are obtained that are expected to simultaneously reproduce both components.

\paragraph{\textbf{Coeval varying mixing model $\text{(Coeval\_VM)}$:}} This model is similar to the coeval fixed mixing model, except that the same set of  $f_{\rm CBM}$ and $\log D_{\rm mix,0}$ for both components is not enforced. In this case, the mixing parameters for both components are independent and are expected to be independently determined by the respective masses of the components under the coeval assumption. For example, CBM is reported to be dependent on the mass \citep[e.g.][]{ClaretandTorres2016, ClaretandTorres2019, Vanrespailleetal2026, Lechienetal2026}. Thus, allowing different mixing parameters for the components clearly captures the effect of  mass on the mixing parameters and shows whether enhanced independent mixing for both components is enough to solve the mass discrepancy in binary systems.

\paragraph{\textbf{Coeval mass-dependent mixing model $\text{(Coeval\_MD)}$:}} Here we assume varying mixing parameters for both components like the coeval varying mixing model except that the values of CBM for both components are not independently free, but coupled to each other by a mass-dependent relation. A common envelope mixing is shared by both components in this  formulation. 

Enhanced mixing has been reported to be required for high-mass stars \citep[e.g.][]{Castroetal2014, HigginsandVink2019, Tkachenkoetal2020} and to partially resolve the mass discrepancy problem \citep{Tkachenkoetal2020}. 
Motivated by the previous observational evidence, and to investigate whether CBM varies systematically with mass in our current sample, we introduced an empirical first-order Taylor-like mass-dependent parameterisation of: 
\begin{equation}
f_{\rm CBM}(M) = f_{\rm CBM,ref} + s_f \left(M - M_{\rm ref}\right)\, ,
\label{eq:fov_mass_law}
\end{equation}
\noindent where $f_{\rm CBM}(M)$ is the CBM parameter assigned to a star of mass $M$, and $f_{\rm CBM,ref}$ is the reference value of CBM parameter evaluated at the reference mass $M_{\rm ref}$. Here,  $M_{\rm ref}$ is chosen as the midpoint of the binary component masses and given as
\begin{equation}
M_{\rm ref} = \frac{M_1 + M_2}{2}\,.
\end{equation}
\noindent We also define
\begin{equation}
s_{f}=\left.\frac{{\rm d}f_{\rm CBM}}{{\rm d}M}\right|_{M_{\rm ref}} 
\end{equation}
\noindent as the local gradient at the reference mass with units of CBM per solar mass ($\mathrm{M}_\odot^{-1}$). A gradient of $s_f=0$ reflects the $\text{coeval\_FM}$ model, where the same mixing parameter values successfully reproduce both components. The choice of reference mass as the mean of the binary masses is to symmetrically parametrise the mixing relation within the binary system.

Because the evolutionary grid is computed only for discrete values of $f_{\rm CBM}$, the inferred values of $f_{\rm CBM} (M)$
are restricted to the available grid points. Consequently, the gradient $s_{\rm f}$ is not interpreted as a precise continuous derivative of CBM with mass. Instead, it is used here as an empirical diagnostic indicating whether the preferred grid-level value of CBM increases, decreases, or remains constant between the two binary components. A finer grid or continuously sampled $f_{\rm CBM}$ would be required to infer a more continuous-like mass dependence.

\section{Results}
\label{sec:results}

\subsection{Influence of Bayesian inferential formulations}
Results for each of the different statistical formulations are obtained.
As expected, their comparison demonstrates that the derived results are systematically influenced by the adopted likelihood constructions. When the dynamical mass is included in the likelihood (like in the PMM formulation), its 
strict demand and comparatively small uncertainty exerts a strong influence on the posterior distribution, resulting in a significantly reduced mass discrepancy. In contrast, when the mass is excluded from the likelihood function (like in PM formulation), the inferred solutions are determined by the combined constraints imposed by the effective temperature, luminosity and surface gravity, allowing the residual mass discrepancy to be evaluated independently.

The Pareto-based model selection procedure further shows that the statistically preferred models are not always those that maximize the evidence or minimise the residual mass discrepancy. Instead, the preferred solutions balance overall statistical support demonstrated by the statistical evidence with agreement between evolutionary and dynamical masses demonstrated by the posterior probability consistent with zero discrepancy. Figure~\ref{fig:pareto} illustrates the Pareto trade-off  between the Bayesian evidence and discrepancy reduction.

\subsection{Physical implications for mixing}
Across the explored grid of evolution models, increasing the values of the mixing parameters generally increases the convective core mass and reduces the inferred evolutionary mass discrepancy, thus corroborating previous findings in the literature \citep[][]{HigginsandVink2019, Tkachenkoetal2020}. However, the dependence is not strictly monotonic. Excessively increasing the values of the mixing parameters may also  shift the evolutionary tracks away from the observed positions in the Hertzsprung–Russell (HR) diagram or fail to satisfy the remaining observational constraints. Consequently, the Bayesian evidence does not invariably favour the strongest available mixing. 

\begin{figure*}[h]
\centering
\includegraphics[width=0.49\linewidth]{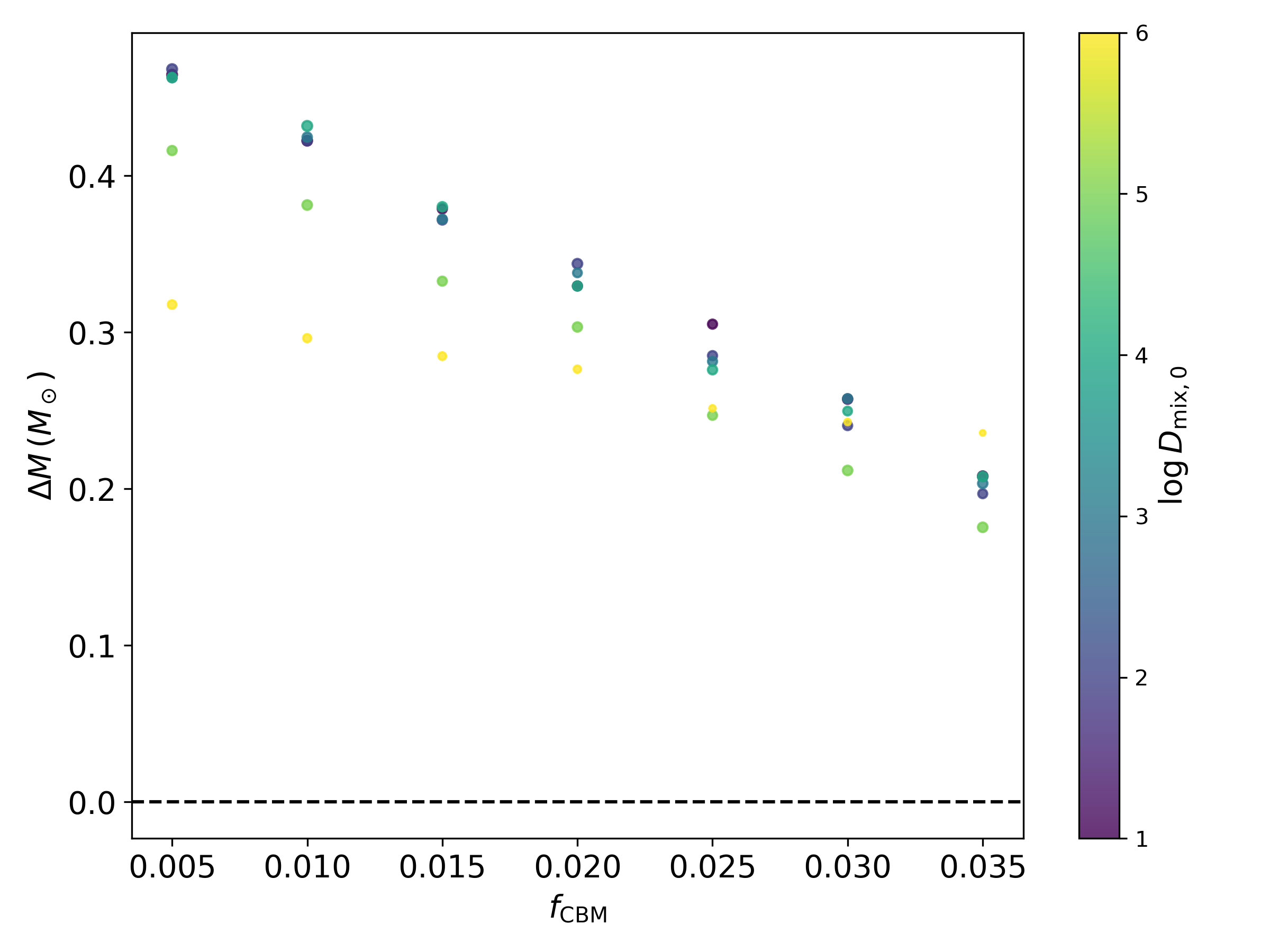}
\includegraphics[width=0.49\linewidth]{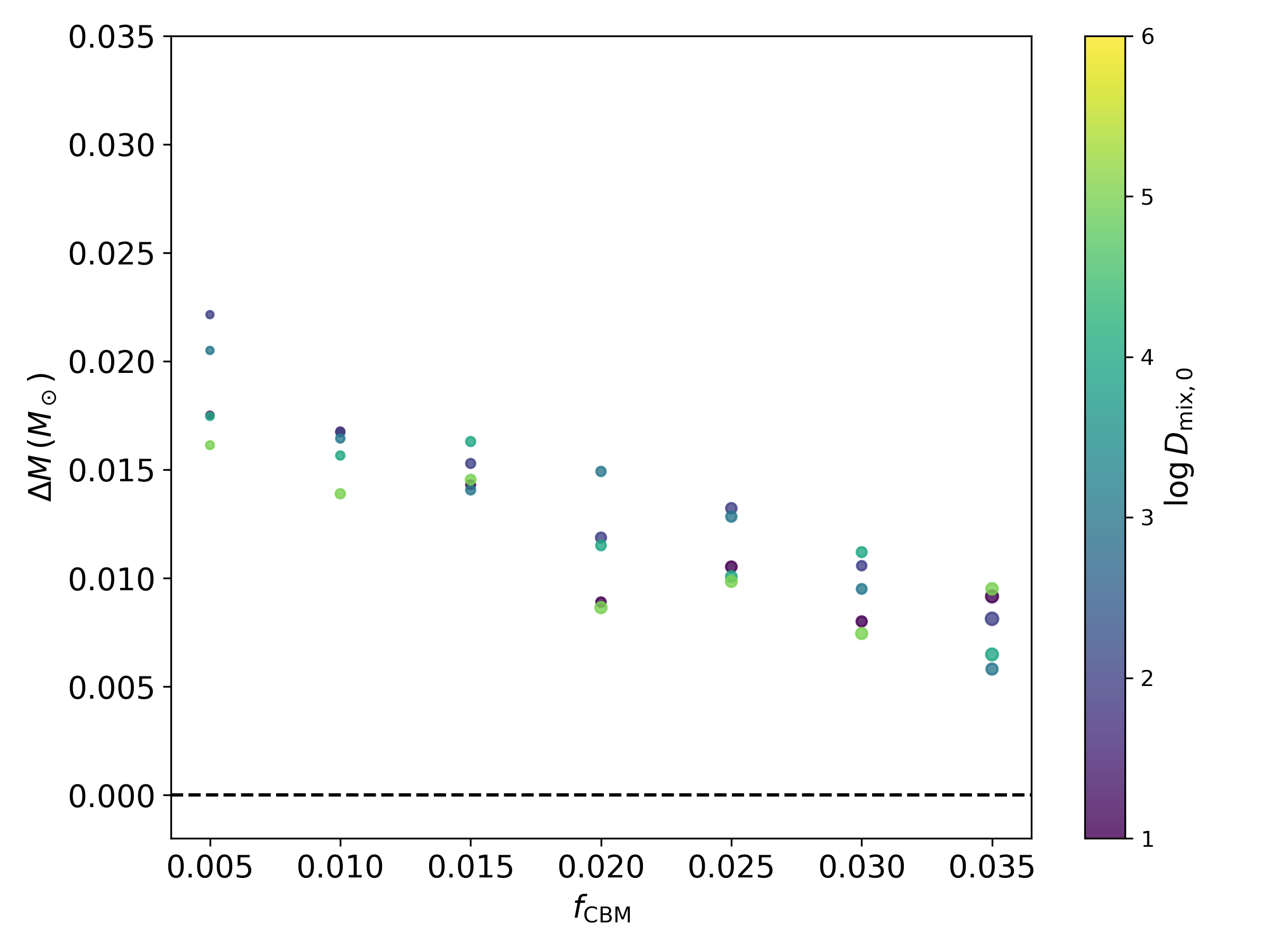}
\caption{The mass discrepancy, $\Delta M$, for different values of CBM, $f_{\rm CBM}$, for the Pareto model (PM; left panel) and mass model (MM; right panel) for the target CW~Cep.}
\label{fig:single_deltaM_mixing}

\end{figure*}

\begin{figure*}[h]
\centering
\includegraphics[width=0.49\linewidth]{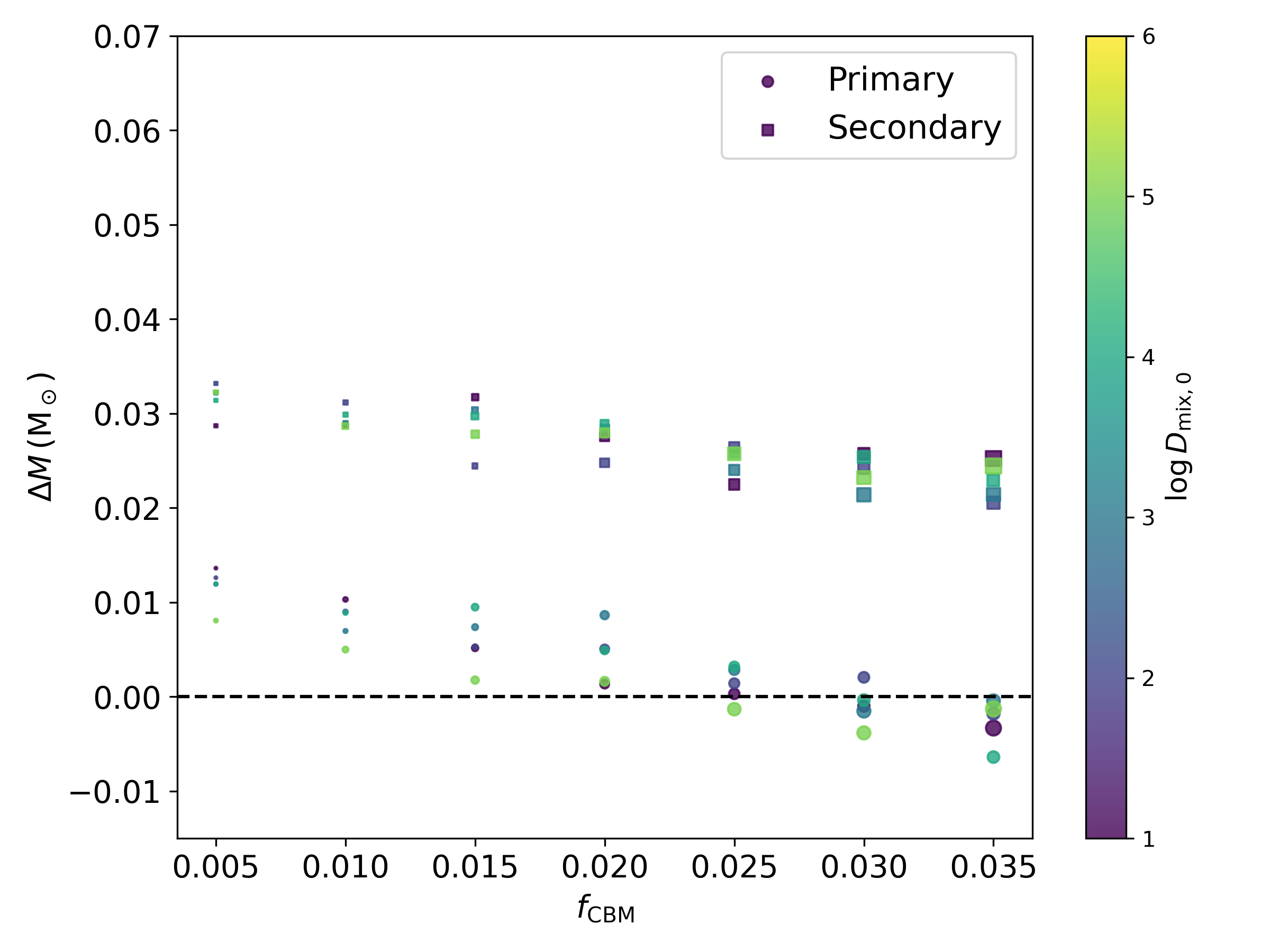}
\includegraphics[width=0.49\linewidth]{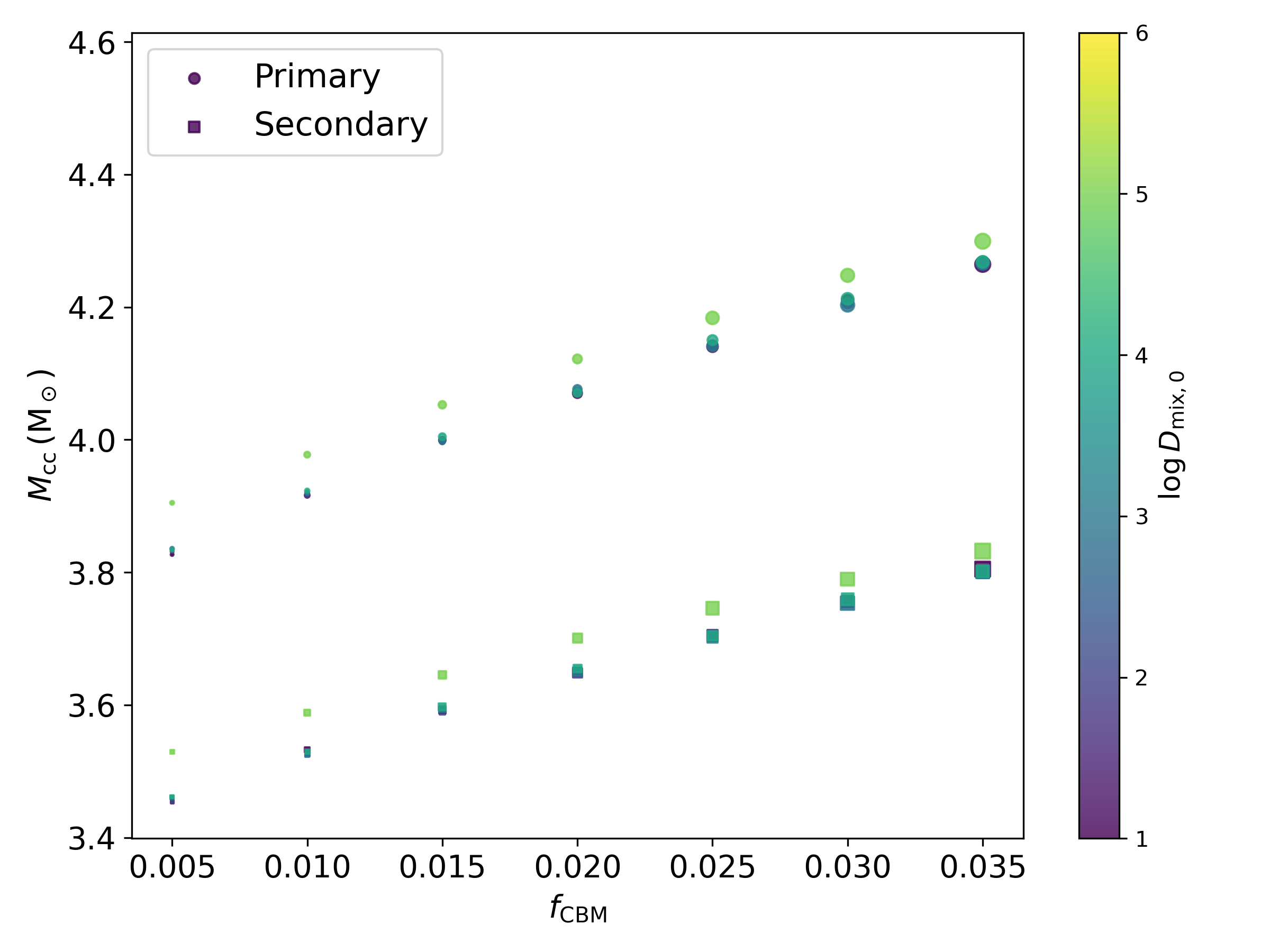}
\caption{The mass discrepancy, $\Delta M$, (left panel) and convective core mass (right panel) for different values of CBM, $f_{\rm CBM}$, for the coeval fixed-mixing model for the target CW~Cep.}\label{fig:coeval_massdis_mcc}
\end{figure*}

As clearly visible in Figs.~\ref{fig:single_deltaM_mixing} and \ref{fig:coeval_massdis_mcc}, which summarise the behaviour of the preferred mixing parameters on the residual mass discrepancy and the corresponding convective-core masses, both CBM and envelope mixing have a significant impact on the measured mass discrepancy and the convective core mass, as already found by \citet{Tkachenkoetal2020}. However, the effect of envelope mixing appears to become comparably stronger as its value increases.

In addition, while the inferred convective core masses correlate with larger values of mixing and anti-correlates with the mass discrepancy, stars with similar core masses do not always exhibit identical discrepancies. This indicates that the inferred stellar properties depend jointly on the adopted mixing parameters, the evolutionary state, and the observational constraints entering the inference. An average of $3-9\%$ of the relative convective core mass growth arises from the increase in the values of the mixing parameters for the chosen input physics in the adopted model grid.

%%% Conny stopped here Sunday night, continuing tomorrow

\subsection{Coeval binary inference}
\begin{figure}[h]
\centering
\includegraphics[width=0.90\linewidth]{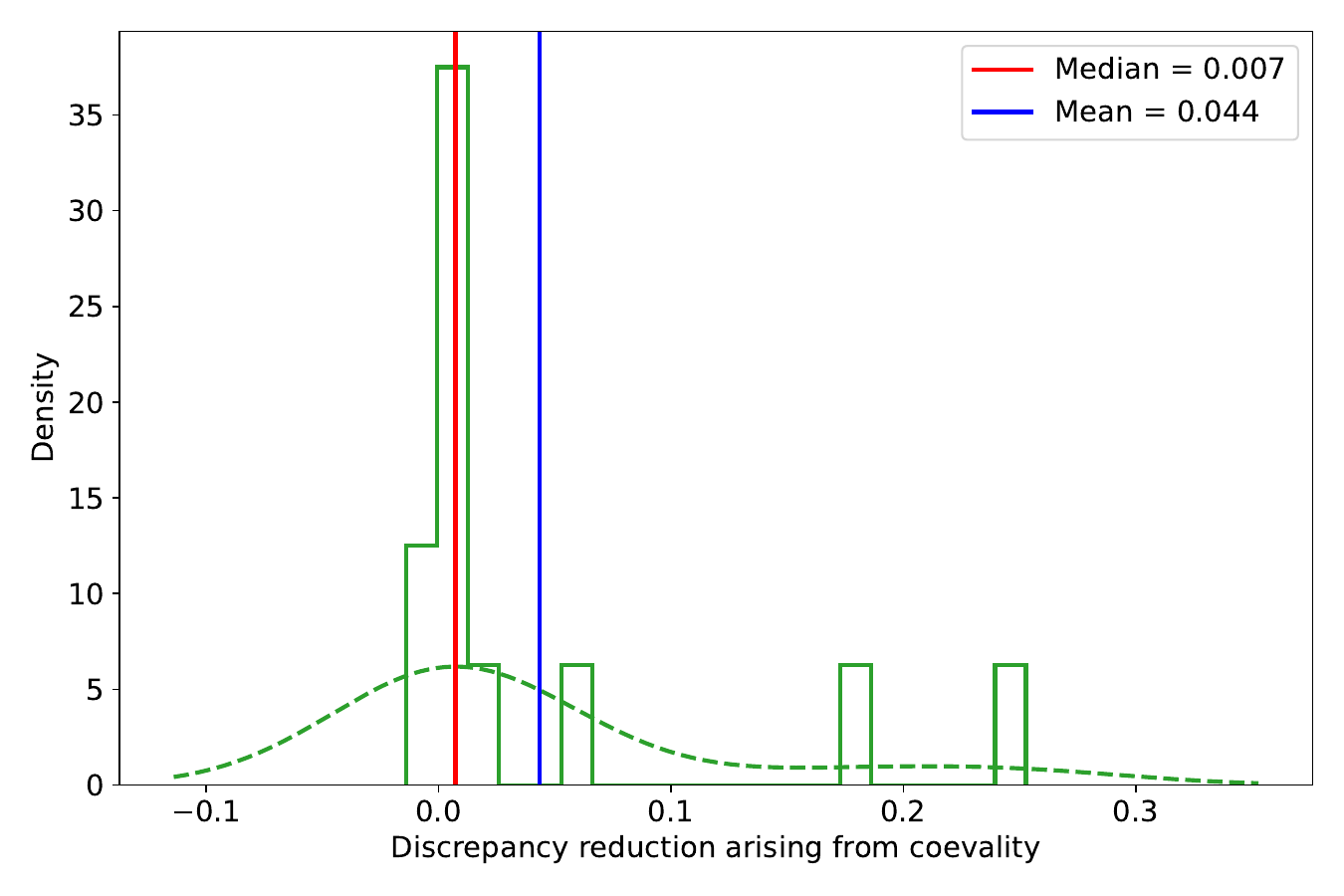}
\caption{Mass discrepancy reduction arising from the enforcement of coevality when PMM and Coeval\_VM are compared.}
\label{fig:discrepancy_reduction_from_coevality.}
\end{figure}

The coeval analyses provide an additional physically motivated constraint because both components of each binary system are required to share a common age. This reduces the available parameter space and therefore provides a more stringent test of the adopted mixing parameters. 
By comparing PMM and Coeval\_VM, which share similar likelihood function conditions and independent mixing parameters for both components, we find that enforcing coeval evolution reduces the observed mass discrepancy by a median of only 0.7\% and an average of $4\%$ for both components. The additional constraint imposed by the common age barely
improves the optimisation procedure. The slight reduction of the mass discrepancy 
for the adopted input physics of the grid is
shown in Fig.\,\ref{fig:discrepancy_reduction_from_coevality.},
where we overplot the Kernel Density Estimate (KDE) of the sparse histogram to place the result into context given our small sample size.
%is overplotted as the dashed lines. 
%KDE is a non-parametric method of estimating the probability density function of a random variable \citep{Rosenblatt1956, Parzen1962, XieandWu2014} and helps to visualise the range and distributions of a random variable. 

For other coeval models such as Coeval\_FM and Coeval\_MD, which share slightly different mixing parameterisation with PMM (although with similar likelihood conditions), the comparisons yielded inconclusive results.
%show predominantly mixed results for improvement (or otherwise) of the agreement of the dynamical masses with the evolutionary masses. This suggests that the impact of assuming coeval evolution on the mass discrepancy becomes more pronounced if the expected enhanced mixing parameter values are already accounted for. 
Comparing the left panel of Fig.~\ref{fig:coeval_massdis_mcc} to Fig.~\ref{fig:single_deltaM_mixing} demonstrates the impact of assuming coeval evolution on the mass discrepancy for the case of CW\,Cep.

The coeval\_FM formulation reproduces several binary systems with a common mixing parameter value for both components fairly well, whereas other systems require different preferred mixing efficiencies for the two components. It is not clear why some components of comparably similar mass difference  between the components
%masses of the binary system 
may be reproduced by similar mixing parameter values while others are not,
as our sample is too small to derive conclusions at population level.

\begin{figure}[h]
\centering
\includegraphics[width=0.90\linewidth]{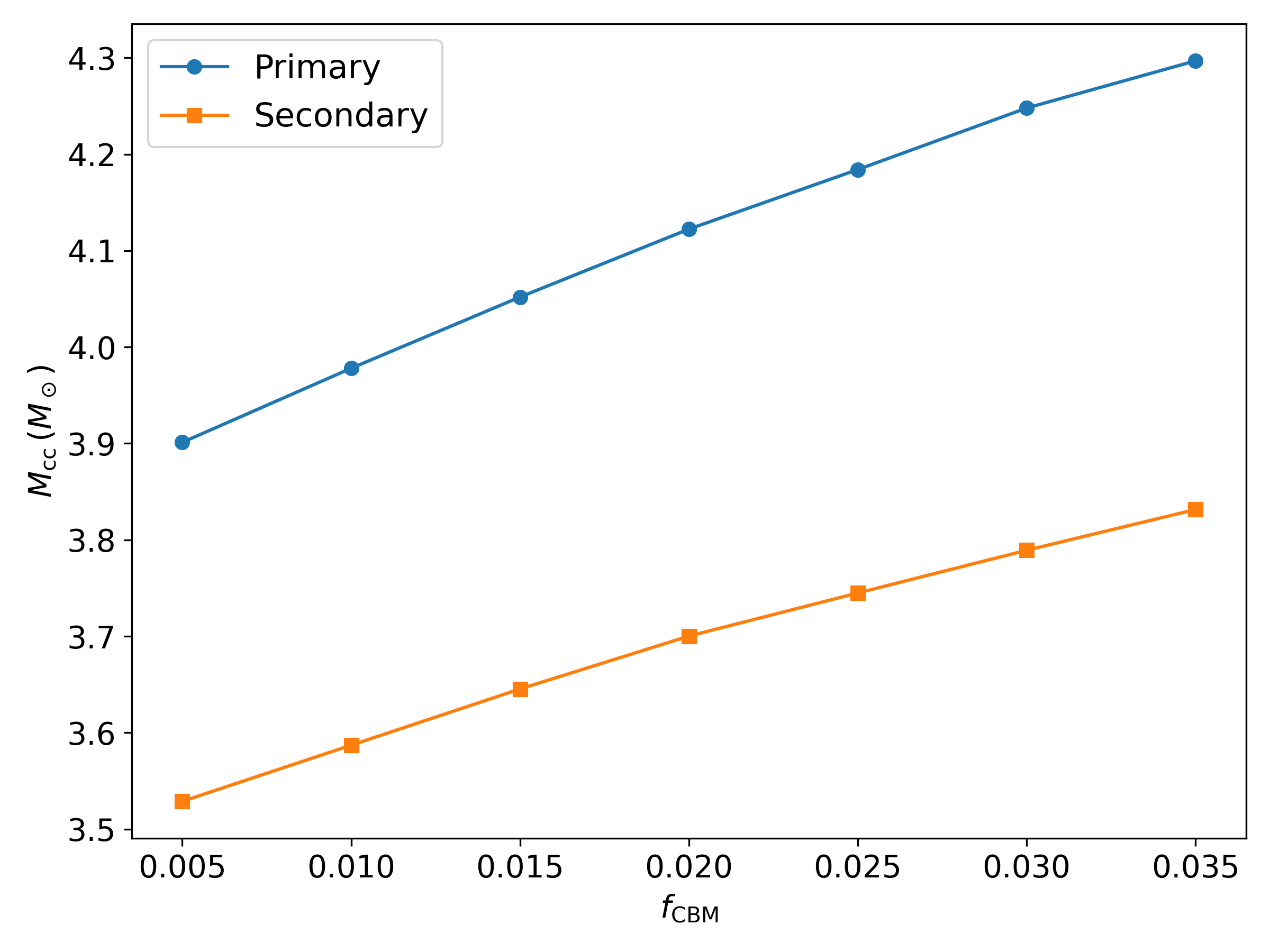}
\caption{Convective core mass as a function of mass-dependent $f_{\rm CBM}$ at the preferred coeval age and envelope mixing strength for the Coeval\_MD mass of CW~Cep.} 
\label{fig:Mcc_mass-dependent_overshooting.}
\end{figure}

Allowing component-dependent mixing (like in the case of Coeval\_VM models) improves the agreement for most binaries in the sample because there are more degrees of freedom.
%, confirming the conclusion that component-based mass-dependent mixing better reproduces the parameters of the binary system.   Additionally, the Coeval\_MD models  also corroborates this conclusion of mass-dependent CBM. 
Figure\,~\ref{fig:Mcc_mass-dependent_overshooting.} shows an example of the convective core mass relation with $f_{\rm CBM}$ at the best fit $\log D_{\rm mix,0}$, coeval age, 
and mass of the Coeval\_MD model for CW~Cep. The figure illustrates the local gradient for this particular binary when we rely on the chosen input physics of the adopted grid.
 
\begin{figure}[h]
\centering
\includegraphics[width=0.90\linewidth]{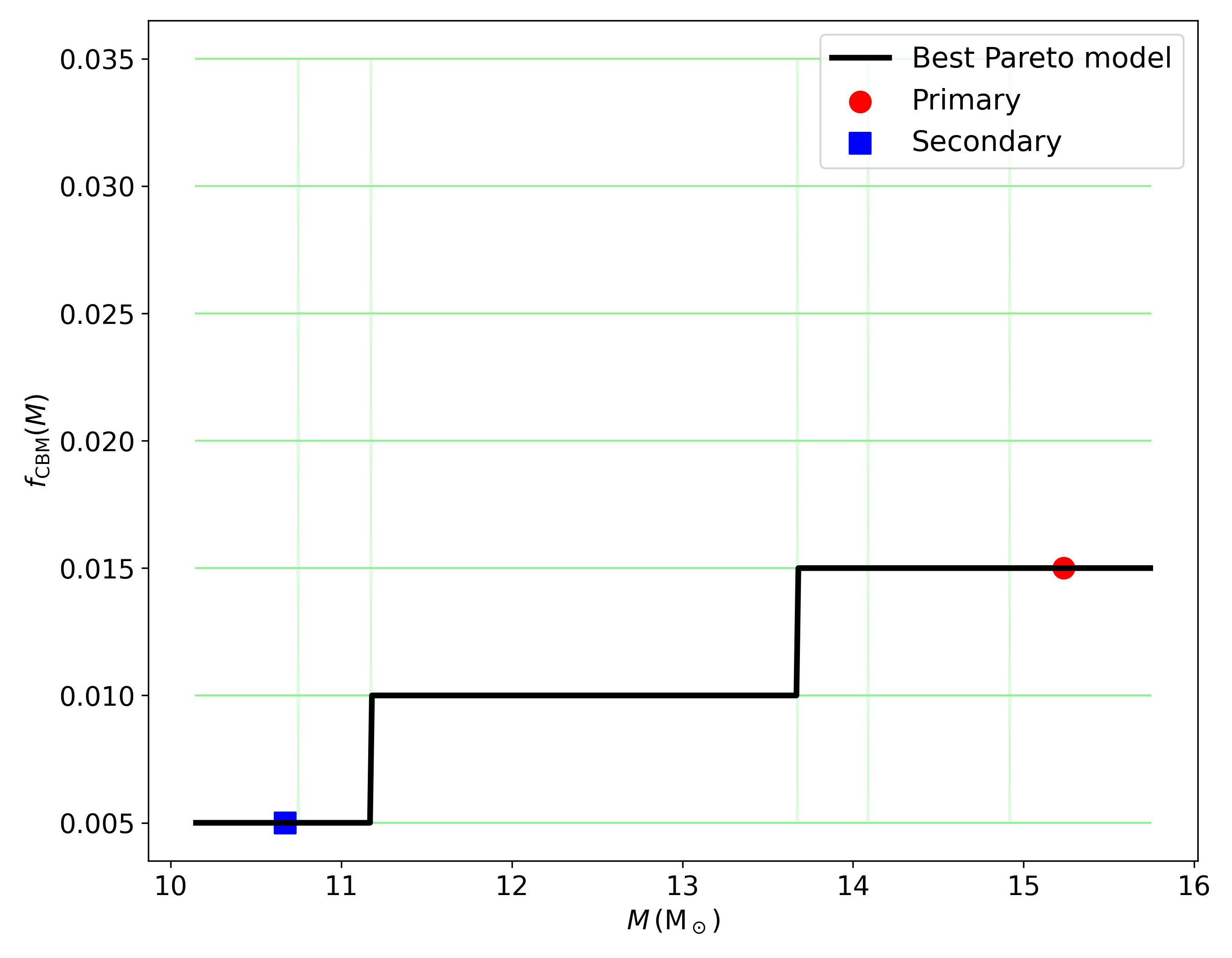}
\caption{Example of the mass-dependent CBM, $f_{\rm CBM}$ versus stellar mass for HD\,254346. The green lines denote the explored CBM values in the grid.
}
\label{fig:Mcc_mass-dependent_overshooting_exploration}
\end{figure}

Figure~\ref{fig:Mcc_mass-dependent_overshooting_exploration} illustrates
the mass-dependent CBM explored from our Bayesian analyses over-plotted with the preferred CBM for both components in the case of HD\,254346.
While the figure suggests that more massive components require more efficient mixing, the apparent step-like behaviour is just a consequence of the discrete sampling of $f_{\rm CBM}$ in the underlying MESA grid rather than evidence for a discontinuous physical relation. The values of $s_{\rm f}$ ($0-2\times10^{-3}~M^{-1}_{\rm \odot}$) obtained for the Coeval\_MD models are only weakly constrained and should only be interpreted qualitatively.

The derived parameters of the targets are listed in 
Table\,\ref{tab:single_case_model} for the single case model and in 
Table\,\ref{tab:coeval_bin_case_model} for the coeval binary case model. We emphasise that systems whose components were not successfully reproduced by a coeval model were also not well reproduced in the single star case, pointing towards other physical processes at play 
that are not captured by the input physics in our MESA model grid.

\begin{table*}[ht]
\begin{center}
\small
\setlength{\tabcolsep}{11pt}
\caption{Derived stellar parameters for the models in the single star case. $\psi\,$~Ori was removed and excluded from further analysis because one of the components gave $X_{\rm c}$ at the edge of the grid, suggesting that it could not be resolved within the current grid limit. RM, MM, PM and PMM denote the reference model, the mass model, the Pareto model and the Pareto mass model,  respectively.}\label{tab:single_case_model}
\begin{tabular}[t]{ccccccc}
\hline\hline
\textbf{RM}\\
\hline
Star   & CW Cep  & HD 254346 & HD 74455 & KX Vel &V453 Cygni&  V1295 Sco  \\
\hline
$M_{1}\, (\mathrm{M}_{\odot})$	& 13.42(29)& 15.89(43) & 14.93(42)&14.13(33) & 17.13(37)& 18.39(36)\\
$M_{2}\, (\mathrm{M}_{\odot})$	& 12.37(21)  &- &9.37(21) &7.74(15) &12.67(32) &13.34(23) \\
$M_\mathrm{cc, 1} \, (\mathrm{M}_{\odot})$ &4.07(15) &4.20(21) &4.95(27) &3.25(15) &4.43(17) & 5.63(19)\\
$M_\mathrm{cc, 2} \, (\mathrm{M}_{\odot})$ & 3.67(11)& - &2.33(13) &2.11(7) &3.62(16) & 4.02(12)\\
$\frac{M_\mathrm{cc, 1}}{M_{1}}$ &0.30  &0.26 &0.33 &0.23 & 0.26&0.31 \\	
$\frac{M_\mathrm{cc, 2}}{M_{2}}$ &0.29 & - &0.25 & 0.27& 0.29& 0.30\\	
$X_{\rm c,1}$ & &0.26(2) &0.53(3) &0.206(22) &0.197(3) &0.316(6) \\	
$X_{\rm c,2}$ & 0.53(1)& - & 0.50(3)&0.655(29) &0.475(9) & 0.501(9) \\	
$\mathrm{age}_1\, $(Myr) &6.2(3)  &8.5(4) &4.70(69) &10.67(44) &8.37(24) &6.54(18) \\
$\mathrm{age}_2\, $(Myr) &6.3(3)  &- &11.5(1.6) &5.3(2.4) &7.52(41) &6.37(29) \\
\hline\hline
\textbf{MM}\\
\hline
Star   & CW Cep  & HD 254346 & HD 74455 & KX Vel & V453 Cygni & V1295 Sco \\
\hline
$M_{1}\, (\mathrm{M}_{\odot})$	&12.959(49)  &15.25(4) &16.29(3) &13.68(3) &14.40(19) & 17.26(2)\\
$M_{2}\, (\mathrm{M}_{\odot})$	&11.89(5)  &- & 10.29(4)&8.25(4) &11.42(14) & 12.67(3)\\
$M_\mathrm{cc, 1} \, (\mathrm{M}_{\odot})$ &4.27(3) & 4.59(8)&5.78(15) &3.59(6) &4.89(9) &6.56(1) \\
$M_\mathrm{cc, 2} \, (\mathrm{M}_{\odot})$ &3.82(2)  &- &2.75(9) &2.22(8) &3.51(7) & 4.16(2)\\
$\frac{M_\mathrm{cc, 1}}{M_{1}}$ &0.33  &0.30 & 0.35& 0.26&0.34 & 0.38\\	
$\frac{M_\mathrm{cc, 2}}{M_{2}}$ & 0.32 & -&0.27 & 0.27&0.31 & 0.33\\	
$X_{\rm c,1}$ &0.529(5)  &0.301(18) &0.556(29) &0.256(17) & 0.294(2)& 0.383(5) \\	
$X_{\rm c, 2}$ & 0.553(5)&- &0.523(35) &0.619(31) &0.503(7) &0.528(7) \\	
$f_{\rm CBM, 1}$ &0.035 &0.005 &0.005 &0.02 &0.035 &0.035 \\ 
$f_{\rm CBM, 2}$ & 0.035&- &0.005 & 0.005&0.035 & 0.035\\ 
$\log D_\mathrm{mix, 1}\, (\mathrm{cm}^{2}\,\mathrm{s}^{-1})$ & 2.0 &6.0 & 3.0 & 2.0&6.0 & 6.0 \\
$\log D_\mathrm{mix, 2}\, (\mathrm{cm}^{2}\,\mathrm{s}^{-1})$ & 5.0 &- &3.0 &2.0 & 5.0& 5.0\\
$\mathrm{age}_1\, $(Myr) &7.46(17) &10.11(29) &3.74(61) & 11.83(26)& 12.96(34)& 8.21(8)\\
$\mathrm{age}_2\, $(Myr) &7.91(23)  &- &8.91(1.3) &7.58(3) & 10.67(41)& 7.98(26)\\
\hline\hline
\textbf{PM}\\
\hline
Star   & CW Cep  & HD 254346 & HD 74455 & KX Vel & V453 Cygni & V1295 Sco \\
\hline
$M_{1}\, (\mathrm{M}_{\odot})$	&	13.13(29)	&	15.27(46)	&	14.95(42)	&	13.65(33)	&	15.60629	&	17.32(36)	\\
$M_{2}\, (\mathrm{M}_{\odot})$	&	12.11(22)	&	-	&	9.37(21)	&	7.74(15)	&	12.29792	&	12.98(25)	\\
$M_\mathrm{cc, 1}\, (\mathrm{M}_{\odot})$	&	4.39(15)	&	4.80(24)	&	4.95(28)	&	3.74(16)	&	5.48(19)	&	6.59(20)	\\
$M_\mathrm{cc, 2}\, (\mathrm{M}_{\odot})$	&	3.93(11)	&		-&	2.33(13)	&	2.11(7)	&	3.94(16)	&	4.32(13)	\\
$\frac{M_\mathrm{cc, 1}}{M_{1}}$	&	0.33	&		0.31&	0.33	&	0.27	&	0.35	&	0.38	\\
$\frac{M_\mathrm{cc, 2}}{M_{2}}$	&		0.32&	-	&	0.25	&	0.27	&	0.32	&	0.33	\\
$X_\mathrm{c, 1}$	&	0.533(5)	&	0.321(17)	&	0.534(29)	&	0.269(18)	&	0.297(2)	&	0.383(4)	\\
$X_\mathrm{c, 2}$	&	0.556(5)	&	-	&	0.501(35)	&	0.655(29)	&	0.510(8)	&	0.531(7)	\\
$f_{\rm CBM, 1}$	&	0.035	&	0.03	&	0.005	&	0.025	&	0.035	&	0.035	\\
$f_{\rm CBM, 2}$	&	0.035	&		-&	0.005	&	0.005	&	0.035	&	0.035	\\
$\log D_\mathrm{mix, 1}\, (\mathrm{cm}^{2}\,\mathrm{s}^{-1})$	&	5	&	4	&	3	&	3	&	6	&	6	\\
$\log D_\mathrm{mix2}\, (\mathrm{cm}^{2}\,\mathrm{s}^{-1})$	&	5	&		-&	1	&	1	&	5	&	5	\\
age$_{1}$ ({\rm Myr})	&	7.36(39)	&	9.76(50)	&	4.70(69)	&	12.08(55)	&	11.21(48)	&	8.16(29)	\\
age$_{2}$ ({\rm Myr})	&	7.55(38)	&	-	&	11.52(1.63)	&	5.34(2.40)	&	9.07(55)	&	7.57(38)	\\
\hline\hline
\textbf{PMM}\\
\hline
Star   & CW Cep  & HD 254346 & HD 74455 & KX Vel & V453 Cygni & V1295 Sco \\
\hline
$M_{1}\, (\mathrm{M}_{\odot})$	&	12.96(5)	&	15.25(4)	&	16.29(3)	&	13.68(3)	&	14.40(19)	&	17.26(2)	\\
$M_{2}\, (\mathrm{M}_{\odot})$	&	11.89(5)	&	-	&	10.29(4)	&	8.25(4)	&	11.42(14)	&	12.673)	\\
$M_\mathrm{cc, 1}\, (\mathrm{M}_{\odot})$	&	4.28(3)	&	4.59(7)	&	5.78(15)	&	3.77(6)	&	4.89(9)	&	6.56(1)	\\
$M_\mathrm{cc, 2}\, (\mathrm{M}_{\odot})$	&	3.78(2)	&	-	&	2.74(11)	&	2.22(8)	&	3.51(7)	&	4.12(2)	\\
$\frac{M_\mathrm{cc, 1}}{M_{1}}$	&	0.33	&	0.30	&	0.35	&	0.28	&	0.34	&		0.38\\
$\frac{M_\mathrm{cc, 2}}{M_{2}}$	&	0.32	&	-	&	0.27	&	0.27	&	0.31	&	0.33	\\
$X_\mathrm{c, 1}$	&	0.529(5)	&	0.301(19)	&	0.556(29)	&	0.269(18)	&	0.294(2)	&	0.383(5)	\\
$X_\mathrm{c, 2}$	&	0.551(5)	&	-	&	0.523(36)	&	0.619(31)	&	0.503(7)	&	0.527(8)	\\
$f_{\rm CBM, 1}$	&	0.035	&	0.005	&	0.005	&	0.025	&	0.035	&	0.035	\\
$f_{\rm CBM, 2}$	&	0.03	&	-	&	0.005	&	0.005	&	0.035	&	0.035	\\
$\log D_\mathrm{mix, 1}\, (\mathrm{cm}^{2}\,\mathrm{s}^{-1})$	&	4	&	6	&	3	&	2	&	6	&	6	\\
$\log D_\mathrm{mix, 2}\, (\mathrm{cm}^{2}\,\mathrm{s}^{-1})$	&	5	&	-	&	2	&	2	&	5	&	2	\\
age$_{1}$ ({\rm Myr})	&	7.47(18)	&	10.11(29)	&	3.74(61)	&	12.02(28)	&	12.96(34)	&	8.21(7)	\\
age$_{2}$ ({\rm Myr})	&	7.76(22)	&	-	&	8.92(1.4)	&	7.39(2.15)	&	10.67(41)	&	7.84(26)	\\
\hline
\end{tabular}
\end{center}
\end{table*}

\begin{table*}[!htbp]
\begin{center}
\small
\setlength{\tabcolsep}{11pt}
\renewcommand{\arraystretch}{1.09}
\caption{Derived stellar parameters for the models in the coeval binary case. Coeval\_FM, coeval\_VM and coeval\_MD denote the coeval fixed mixing model, the coeval varying mixing model and the coeval mass-dependent mixing model, respectively.}\label{tab:coeval_bin_case_model}
\begin{tabular}[t]{ccccccc}
\hline\hline
\textbf{Coeval\_FM}\\
\hline\hline
Star   & CW Cep  & HD 254346 & HD 74455 & KX Vel & V453 Cygni & V1295 Sco \\
\hline
$M_{1}\, (\mathrm{M}_{\odot})$	&	12.949(53)	&	15.24(4)	&	16.29(3)	&	13.68(3)	&	14.79(17)	&	17.26(2)	\\
$M_{2}\, (\mathrm{M}_{\odot})$	&	11.901(46)	&	10.67(5)*	&	10.30(4)	&	8.25(4)	&	11.26(11)	&	12.67(3)	\\
$M_\mathrm{cc, 1}\, (\mathrm{M}_{\odot})$	&	4.29(7)	&	3.69(8)	&	5.66(13)	&	3.10(7)	&	4.60(90	&	6.18(2)	\\
$M_\mathrm{cc, 2}\, (\mathrm{M}_{\odot})$	&	3.83(13)	&	2.79(2)	&	3.08(4)	&	2.09(1)	&	3.42(5)	&	4.17(1)	\\
$\frac{M_\mathrm{cc, 1}}{M_{1}}$	&	0.33	&	0.24	&	0.35	&		0.23&	0.31	&		0.36\\
$\frac{M_\mathrm{cc, 2}}{M_{2}}$	&	0.32	&	0.26	&	0.30	&	0.25	&	0.30	&	0.33	\\
$X_\mathrm{c, 1}$	&	0.528(9)	&	0.205(19)	&	0.532(26)	&	0.211(21)	&	0.280(2)	&	0.371(4)	\\
$X_\mathrm{c, 2}$	&	0.558(9)	&	0.484(7)	&	0.632(11)	&	0.565(4)	&	0.495(5)	&	0.534(2)	\\
$f_\mathrm{\rm CBM}$	&	0.035	&	0.005	&	0.005	&	0.005	&	0.035	&	0.035	\\
$\log D_\mathrm{mix,0}\, (\mathrm{cm}^{2}\,\mathrm{s}^{-1})$	&	5	&	2	&	1	&	4	&	5	&	5	\\
age (Myr)	&	7.69(15)	&	9.64(20)	&	4.24(51)	&	11.14(26)	&	11.29(21)	&	7.76(7)	\\
\hline\hline
\textbf{Coeval\_VM}\\
\hline
Star   & CW Cep & HD 254346 & HD 74455 & KX Vel & V453 Cygni & V1295 Sco \\
\hline
$M_{1}\, (\mathrm{M}_{\odot})$	&	12.957(46)	&	15.23(4)	&	16.29(3)	&	13.68(29)	&	14.79(16)	&	17.26(2)	\\
$M_{2}\, (\mathrm{M}_{\odot})$	&	11.891(46)	&	10.68(5)*	&	10.30(4)	&	8.25(4)	&	11.28(11)	&	12.66(3)	\\
$M_\mathrm{cc, 1}\, (\mathrm{M}_{\odot})$	&	4.27(2)	&	4.78(5)	&	5.96(9)	&	4.06(5)	&	4.55(8)	&	6.16(2)	\\
$M_\mathrm{cc, 2}\, (\mathrm{M}_{\odot})$	&	3.69(2)	&	2.69(2)	&	3.09(4)	&	2.03(1)	&	3.44(5)	&	4.05(1)	\\
$\frac{M_\mathrm{cc, 1}}{M_{1}}$	&	0.33	&	0.31	&	0.37	&	0.29	&	0.31	&	0.36	\\
$\frac{M_\mathrm{cc, 2}}{M_{2}}$	&	0.31	&	0.25	&	0.30	&	0.25	&	0.30	&	0.32	\\
$X_\mathrm{c, 1}$	&	0.529(4)	&	0.284(1)	&	0.551(25)	&	0.297(16)	&	0.278(2)	&	0.371(5)	\\
$X_\mathrm{c, 2}$	&	0.546(4)	&	0.452(7)	&	0.630(11)	&	0.543(5)	&	0.497(5)	&	0.522(2)	\\
$f_{\rm CBM, 1}$	&	0.035	&	0.035	&	0.02	&	0.035	&	0.035	&	0.035	\\
$f_{\rm CBM, 2}$	&	0.025	&	0.005	&	0.005	&	0.005	&	0.035	&	0.025	\\
$\log D_\mathrm{mix, 1}\, (\mathrm{cm}^{2}\,\mathrm{s}^{-1})$	&	2	&	4	&	2	&	2	&	3	&	2	\\
$\log D_\mathrm{mix, 2}\, (\mathrm{cm}^{2}\,\mathrm{s}^{-1})$	&	2	&	3	&	2	&	1	&	5	&	5	\\
age (Myr)	&	7.45(14)	&	10.62(21)	&	4.30(56)	&	12.27(28)	&	11.18(19)	&	7.69(7)	\\
\hline\hline
\textbf{Coeval\_MD}\\
\hline
Star   & CW Cep  & HD 254346 & HD 74455 & KX Vel & V453 Cygni & V1295 Sco \\
\hline
$M_{1}\, (\mathrm{M}_{\odot})$	&	12.945(49)	&	15.23(4)	&	16.29(3)	&	13.68(3)	&	14.78(16)	&	17.26(2)	\\
$M_{2}\, (\mathrm{M}_{\odot})$	&	11.898(46)	&	10.67(5)*	&	10.30(4)	&	8.25(4)	&	11.25(11)	&	12.67(3)	\\
$M_\mathrm{cc, 1}\, (\mathrm{M}_{\odot})$	&	4.29(3)	&	4.10(6)	&	5.77(11)	&	3.29(7)	&	4.59(8)	&	6.18(2)	\\
$M_\mathrm{cc, 2}\, (\mathrm{M}_{\odot})$	&	3.83(2)	&	2.75(2)	&	3.08(4)	&	2.07(1)	&	3.42(4)	&	4.17(1)	\\
$\frac{M_\mathrm{cc, 1}}{M_{1}}$	&	0.33	&	0.27	&	0.35	&	0.24	&	0.31	&		0.36\\
$\frac{M_\mathrm{cc, 2}}{M_{2}}$	&	0.32	&	0.26	&	0.30	&	0.25	&	0.30	&	0.33	\\
$X_\mathrm{c, 1}$	&	0.529(4)	&	0.240(17)	&	0.539(25)	&	0.232(19)	&	0.280(2)	&	0.371(5)	\\
$X_\mathrm{c, 2}$	&	0.558(3)	&	0.474(7)	&	0.631(11)	&	0.559(5)	&	0.495(5)	&	0.534(2)	\\
$f_{\rm CBM, 1}$	&	0.035	&	0.015	&	0.01	&	0.01	&	0.035	&	0.035	\\
$f_{\rm CBM, 2}$	&	0.035	&	0.005	&	0.005	&	0.005	&	0.035	&	0.035	\\
$\log D_\mathrm{mix,0}\, (\mathrm{cm}^{2}\,\mathrm{s}^{-1})$	&	5	&	3	&	1	&	2	&	5	&	5	\\
age (Myr)	&	7.69(14)	&	9.97(21)	&	4.26(52)	&	11.31(26)	&	11.30(19)	&	7.77(7)	\\
\hline\hline
\end{tabular}
\end{center}
\end{table*}

\subsection{Separating methodological and physical contributions}
We now compare the results from the various Bayesian fitting procedures (14 different models)
in an attempt to distinguish results arising from the statistical formulation (7 different likelihood functions, for each of the single and binary treatment of the components)
from those associated with different values of mixing in the adopted MESA model grid.
Because the analyses of the likelihood functions
differ primarily in the treatment of the dynamical mass and the demand of equal age (or not), their comparison may provide information on the methodological contribution to the inferred mass discrepancy for the input physics and the free parameter ranges adopted in the MESA grid.

In order to compare the 14 models we define
\begin{equation}
\eta =
1 -
\frac{\left|\Delta M_{\rm ehmix}\right|}
     {\left|\Delta M_\mathrm{min}\right|},
     \label{Eq:efficiency}
\end{equation}
where $\Delta M_{\rm ehmix}$ is the discrepancy between the dynamical mass and the mass of the selected best model for the case of enhanced mixing. For $\Delta M_{\rm min}$ we consider two options:
(i) we take the discrepancy between the dynamical mass and the evolutionary mass of the reference models with the absolute minimal amount of core overshooting and envelope mixing in the grid. Hence for this case, $f_{\rm CBM}=0.005$ and $\log D_{\rm mix,0}=1.0$ are fixed values and $\Delta M_{\rm min}$ only changes due to the variations in the likelihood functions;
(ii) rather than using the RM as a common reference, \(\Delta M_{\min}\) is the mass discrepancy between the dynamical mass and the evolutionary mass obtained from the other model configurations, evaluated at the minimum mixing parameters, \(f_{\rm CBM}=0.005\) and \(\log D_{\rm mix,0}=1.0\), while retaining the same likelihood formulation as adopted for their corresponding enhanced-mixing solutions.
In the case of (i), $\eta$ 
indicates the need of extra mixing compared to models without it 
to resolve the mass discrepancy, where zero means no need for extra mixing. 
For option (ii), \(\eta\) quantifies the extent to which enhanced mixing reduces the mass discrepancy across the different model configurations.
Both options use all the variants of the likelihood functions considered in this work. 

\begin{figure*}%[ht]
\centering
\includegraphics[width=0.49\linewidth]{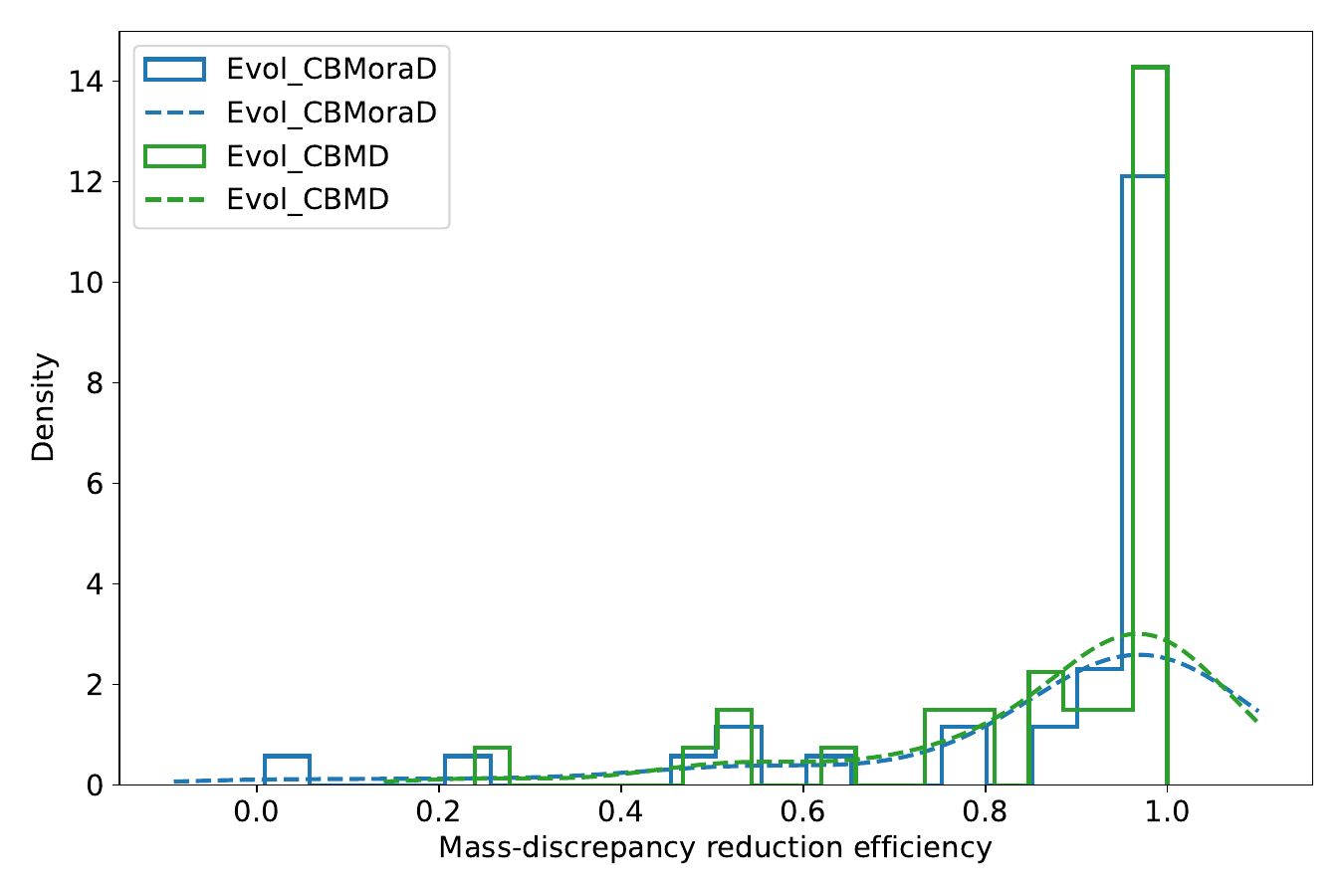}
\includegraphics[width=0.49\linewidth]{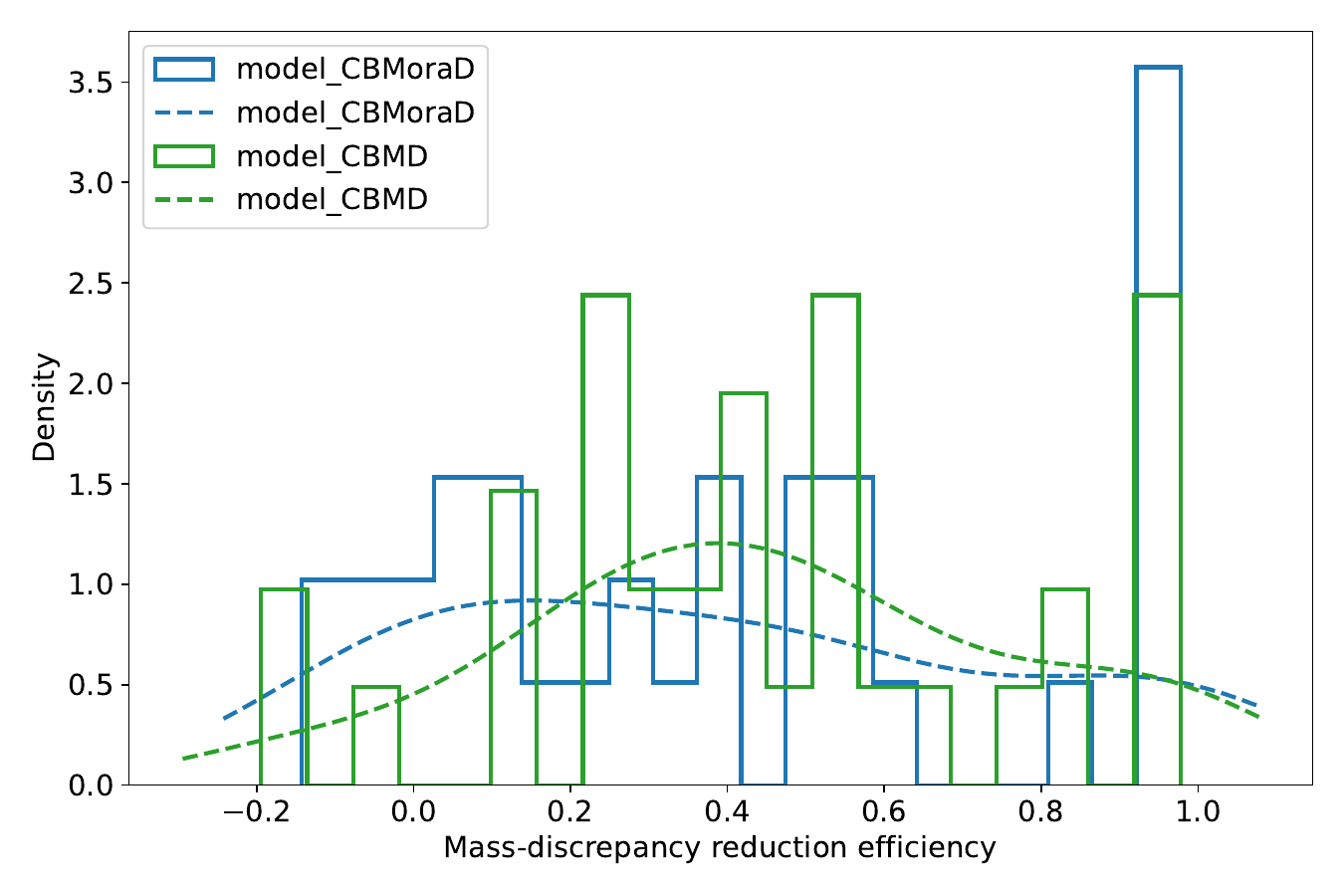}
\caption{
Mass discrepancy reduction efficiency $\eta$ for the cases (i) (left panel) and (ii) (right panel).
All models (i) have $f_{\rm CBM}=0.005$ and $\log D_{\rm mix,0}=1.0$ for their 
$\Delta M_{\rm min}$  in the definition of $\eta$ in Eq.\,(\ref{Eq:efficiency}), while for case (ii), \(\Delta M_{\min}\) is obtained from the corresponding model configurations evaluated at the same minimum mixing parameters (\(f_{\rm CBM}=0.005\) and \(\log D_{\rm mix,0}=1.0\)) under likelihood conditions comparable to those of their corresponding enhanced-mixing solutions. The label
'CBMoraD' denotes preferred models with at least one of $f_{\rm CBM}$ or $\log D_{\rm mix,0}$  above 0.005 or 1, respectively. 'CBMD' represents models for which 
both mixing parameters are greater than their minimum values in the grid. KDEs of the distributions are overplotted in dashed lines. See text for further intepretation.
}\label{fig:mass_discrepancy_reduction_efficiency}
\end{figure*}

The outcomes of $\eta$ in these two cases
are shown in Fig.~\ref{fig:mass_discrepancy_reduction_efficiency}.  
For case (i) in the left panel we find overwhelming evidence that extra mixing is required, irrespective of the chosen likelihood formulation. The
median efficiency of the mass discrepancy reduction $\eta$ is 97\% for both the cases with only core overshooting or envelope mixing enhanced ('Evol\_CBMoraD'), or both types of mixing together
('Evol\_CBMD'). This is in agreement with conclusions in the literature presented in the introductory section of our paper.
For case (ii) the KDEs shift to median $\eta$ values of 42\% for 'model\_CBMoraD' and 37\% for 'model\_CBMD' models. For case (i), $\eta$ is affected by both the enhanced mixing parameters and the variety in the likelihood functions, while for case (ii) the difference mainly comes from the mixing parameters alone, since the variety in the likelihood treatments should affect the numerator and denominator of $\eta$ in a similar way.
Comparing both scenarios hints at approximately 
half of the apparent reduction in the inferred mass discrepancy is attributable 
to the need of extra mixing and half to the methods adopted in
the construction of the likelihood functions, when considering the chosen input physics and parameter ranges of our grid of MESA modes and while evaluating our small sample of eclipsing binaries. 

\section{Discussion and conclusions}\label{sec:discussion}

One of the principal outcomes of this study is that the inferred mass discrepancy depends not only on the adopted stellar physics and the range of free parameters associated with it but also on the construction of the statistical inference problem. By comparing otherwise identical analyses with and without dynamical masses included in the likelihood function, we explicitly isolate the methodological contribution to the inferred discrepancy. This demonstrates that likelihood construction constitutes a source of systematic uncertainty that should be considered when fitting  stellar evolution models.
This conclusion is complementary to the one by \citet{Tkachenkoetal2020}, who stressed the importance of the spectroscopic parameter estimation from atmosphere models, where  
variations in the microturbulent velocity and/or turbulent pressure are often ignored in the likelihood functions.

%\subsection{Influence of inferential formulation on the inferred mass discrepancy}

The mass discrepancy has traditionally been associated with inadequacies in the adopted physics of stellar interiors, with enhanced mixing shown to reduce its magnitude in several studies \citep[e.g.][]{HigginsandVink2019, Tkachenkoetal2020, Burssensetal2023,Lechienetal2026}. Consequently, discussion of the mass discrepancy has largely focused on identifying the missing stellar physics.
Our results suggest that this interpretation may be incomplete. By comparing alternative Bayesian inferential formulations applied to the same stellar evolutionary grid, we find that part of the inferred reduction in the mass discrepancy depends on the statistical formulation of the inference problem itself.  A precisely measured mass in the likelihood comparable to other observables with larger relative uncertainties 
%tends to bias the result towards itself, thereby 
may reduce the inferred discrepancy independently of changes in the underlying stellar physics.
%This result does not diminish the importance of enhanced mixing. It rather 
Our results reinforce the need to separate methodological from physical contributions
%. Rather, it suggests that differences observed in previous studies about 
to resolve the mass discrepancy.

Despite our conclusions, 
mixing remains one of the principal uncertainties in the evolution of intermediate- and high-mass stars. We stress that we have only used one type of mixing profile to describe core overshooting and to mimic envelope mixing, respectively, among many available choices for these two mixing phenomena \citep[cf.\ ][for a broader discussion when performing asteroseismic modelling]{Aerts2021}.
CBM inferred from eclipsing binaries and gravity-mode asteroseismology exhibits substantial star-to-star variation, with reported values depending on the adopted stellar models, evolutionary diagnostics, and fitting methodology \citep[e.g.][]{ClaretandTorres2016, ClaretandTorres2017, ClaretandTorres2018, Mombarg2019,Bowman2020, Pedersenetal2021, Johnston2021b, Vanrespailleetal2026}. Whether this variation reflects genuine physical diversity or limitations of current stellar models remains an open question.

\begin{figure*}[h]
\centering
\includegraphics[width=0.85\linewidth]{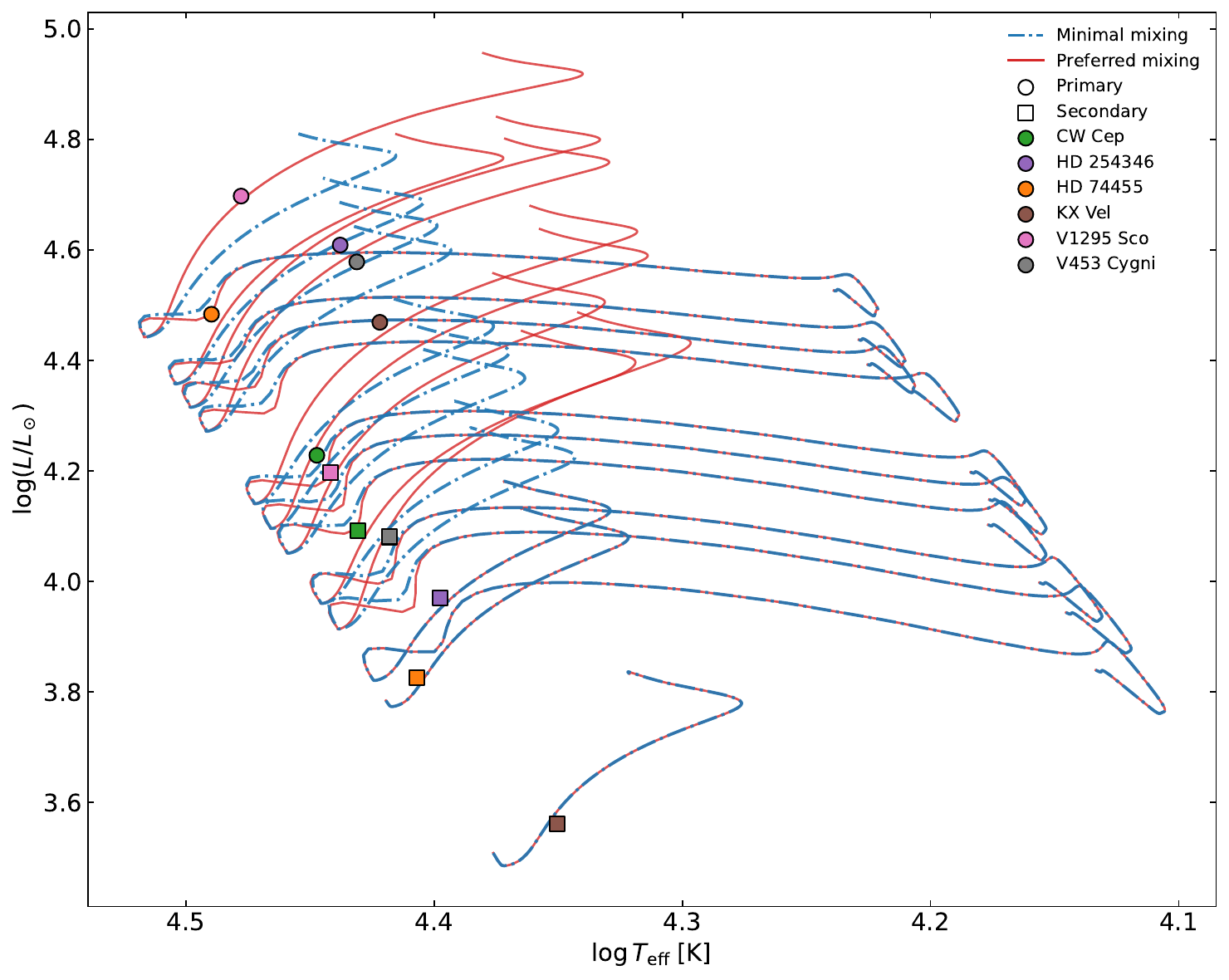}
\caption{HR-diagram showing evolutionary tracks of our binary targets with the statistically preferred mixing (red) and the minimal mixing in the adopted MESA grid of stellar models (blue). 
}
\label{fig:HRD.}
\end{figure*}

Our analysis supports the growing evidence that enhanced mixing reduces the evolutionary mass discrepancy.
This is expected as stronger CBM increases the convective core mass and thus prolongs the main-sequence lifetime and increases the luminosity of a star of a given mass at a given metallicity \citep{Maeder2009, Brottetal2011, Johnston2021b}. Consequently, models with stronger mixing can reproduce the observed positions of the star in the HR~diagram using lower evolutionary masses, thereby bringing them into closer agreement with the independently measured dynamical masses. In addition, increasing mixing parameters also alters the effective temperature, surface gravity, and main-sequence lifetime of stellar models \citep{Aerts2015, Bowman2020}. This is also corroborated by the findings in this work as shown in the HR diagram in Fig.\,\ref{fig:HRD.}, where the evolutionary tracks of both the statistically preferred mixing and the minimal mixing under the coeval\_VM formulation are plotted for the sample.  However, mixing parameters cannot be infinitely increased as excessively large mixing parameters shift the evolutionary tracks away from the observed positions in the HR diagram or require implausible common ages to reproduce both binary components simultaneously. 

In addition, the observed mass discrepancy anti-correlates with the convective core mass corroborating the long established fact in the literature that underestimation of the core mass results in more mass discrepancy.
The coeval binary analyses reinforce these conclusions. Several systems are also satisfactorily reproduced using common mixing parameter values for both components, whereas others favour different mixing efficiencies for the two components. Similar star-to-star variations have been reported in binary and asteroseismic studies \citep[e.g.][]{ClaretandTorres2016, Tkachenkoetal2020}, indicating that mixing cannot be well modelled by assuming that the parameters describing them do not change along the evolution. Results by \citet{Mombarg2019} also show that the shrinking convective cores of intermediate-mass stars can nowadays be measured adequately along the entire main sequence from asteroseismic observables. Keeping this in mind, it is 
unlikely that the internal mixing can be described by a single universal time-independent parameter \citep[cf.\ ][for  the same conclusion from numerical simulations of envelope wave mixing as a function of age]{Vanon2023}. Our empirical mass-dependent CBM follows the same qualitative trend of higher CMB for higher masses, but should be regarded as exploratory because of the small sample size and limited evolutionary stages of our considered binaries.
%discrete sampling of the current evolutionary grid.

Our results further show that models with the highest Bayesian evidence do not always correspond to those with the smallest $\Delta M$, especially if the mass is not enforced. This demonstrates that resolving the mass discrepancy alone is insufficient as successful models are expected to simultaneously reproduce the observed temperatures, luminosities, surface gravities, and coevality constraints. The Pareto-optimal solutions introduced here represent a balance between statistical fit quality and reduction in mass-discrepancy. This corroborates the growing evidence that the best stellar models are not necessarily those with the strongest amount of mixing, but rather those that achieve a physically self-consistent compromise among multiple observables \citep{ClaretandTorres2016, Aerts2021}.

%\subsection{Broader implications}
Bayesian inference has become an increasingly important tool for interpreting modern stellar observations. This is because it naturally combines heterogeneous observables with physically motivated prior information while providing rigorous uncertainty estimates. Most applications, however, employ a single likelihood formulation and implicitly assume that the inferred stellar parameters are insensitive to the choice of observational constraints entering the likelihood. 
The present analysis shows that this assumption may not always be appropriate. The alternative inferential formulations explored here to study the mass discrepancy produce systematically different posterior distributions of the discrepancy despite employing 
just one grid of stellar evolutionary models. Future Bayesian analyses involving 
a variety of stellar evolution model grids with different input physics will surely benefit from explicitly examining the sensitivity of assumed and inferred stellar parameters to the adopted likelihood formulations.

%______________________________________________________________

%\section{Conclusions}
%\label{sec:conclusions}

This study demonstrates that the inferred evolutionary mass discrepancy depends not only on stellar evolution physics but also on the statistical formulation of the inference problem. 
%Different likelihood constructions lead to systematically different posterior distributions and consequently different inferred mass discrepancies. This highlights the importance of carefully distinguishing methodological effects from genuine physical improvements when fitting stellar evolution models.
%Within these inferential strategies, enhanced  mixing remains an important physical mechanism for reducing the evolutionary mass discrepancy. However, 
Comparison of the alternative likelihood formulations indicates that 
%approximately 63-59\%  
about half of the apparent reduction in mass discrepancy may originate from the statistical treatment of the observational constraints rather than solely from changes in the mixing parameters adopted in the used grid of stellar models. The resulting central hydrogen mass fractions for the stars in the coeval cases, ranging from 0.632(11) to 0.205(19), show that the sample spans the early to late main-sequence evolutionary stages. Their coeval ages range from 4.25(51) to 12.27(28) Myr and are determined with precisions of $2$ -- $13\%$ for the adopted input physics.
The methodology presented here could be extended to larger samples and to future analyses incorporating rotation, binary interaction, asteroseismic constraints, or more sophisticated stellar evolution calculations with time-dependent mixing parameters.

\begin{acknowledgements}
This work was supported by the Polish National Science
Foundation (NCN) under grant nr. 2021/43/B/ST9/02972 led by GH.  
DMB gratefully acknowledges UK Research and Innovation (UKRI) in the form of a Frontier Research grant under the UK government's ERC Horizon Europe funding guarantee (SYMPHONY; PI Bowman; grant number: EP/Y031059/1), and a Royal Society University Research Fellowship (PI Bowman; grant number: URF{\textbackslash}R1{\textbackslash}231631).
MV and CA acknowledge financial support from the European Research Council (ERC) under the Horizon Europe programme (Synergy Grant agreement N$^{\circ}$101071505: 4D-STAR). While partially funded by the European Union, views and opinions expressed are however those of the authors only and do not necessarily reflect those of the European Union or the European Research Council. Neither the European Union nor the granting authority can be held responsible for them. 
CA further acknowledges financial support from the Flemish Government under the
long-term structural Methusalem funding program by means of the project SOUL:
Stellar evOlution in fUll gLory, grant METH/24/012 at KU Leuven, Belgium. 

\end{acknowledgements}

%-------------------------------------------------------------------
%\bibliography{references}{}
%\bibliographystyle{aasjournal}

\bibliographystyle{aa} % style aa.bst
\bibliography{bibliography} 

\end{document}